\documentstyle[12pt,axodraw]{article}
\newcommand{\la}[1]{\label{#1}}

\newcommand{\be}{\begin{equation}}
\newcommand{\ee}{\end{equation}}
\newcommand{\ba}{\begin{eqnarray}}
\newcommand{\ea}{\end{eqnarray}}
\newcommand{\bi}{\begin{itemize}}
\newcommand{\ei}{\end{itemize}}
\newcommand{\rmi}[1]{{\mbox{\scriptsize #1}}}
\newcommand{\fig}{Fig.~}
\newcommand{\eq}{Eq.~}
\newcommand{\se}{Sec.~}
\newcommand{\eqs}{Eqs.~}
\newcommand{\nr}[1]{(\ref{#1})}
\newcommand{\tr}{{\rm Tr\,}}
\newcommand{\nn}{\nonumber \\}
\newcommand{\fr}[2]{{\frac{#1}{#2}}}
\newcommand{\msbar}{\overline{\mbox{\rm MS}}}
\newcommand{\lambdamsbar}{\Lambda_{\overline{\rm MS}}}

\newcommand{\bfx}{{\bf x}}

\newcommand{\trasq}{\langle{\rm Tr}A_0^2\rangle}

\newcommand{\ttwo}{\tr A_0^2}
\newcommand{\tthree}{\tr A_0^3}

\newcommand{\ttwob}{\tr B^2}
\newcommand{\tthreeb}{\tr B^3}
\newcommand{\tfourb}{\tr B^4}
\def\lsi{\raise0.3ex\hbox{$<$\kern-0.75em\raise-1.1ex\hbox{$\sim$}}}
\def\gsi{\raise0.3ex\hbox{$>$\kern-0.75em\raise-1.1ex\hbox{$\sim$}}}

\newcommand{\gsim}{\mathop{\gsi}}

\begin{document}

\begin{titlepage}
\begin{flushright}
CERN-TH/98-350\\
NORDITA-98/66HE\\
hep-lat/9811004 
\end{flushright}
\begin{centering}
\vfill

{\bf THE PHASE DIAGRAM OF THREE-DIMENSIONAL\\ 
SU(3)+ADJOINT HIGGS THEORY}
\vspace{0.8cm}

K. Kajantie$^{\rm a,}$\footnote{keijo.kajantie@helsinki.fi},
M. Laine$^{\rm a,b,}$\footnote{mikko.laine@cern.ch},
A. Rajantie$^{\rm c,}$\footnote{a.k.rajantie@sussex.ac.uk}, 
K. Rummukainen$^{\rm d,}$\footnote{kari@nordita.dk} and
M. Tsypin$^{\rm e,}$\footnote{tsypin@td.lpi.ac.ru} \\

\vspace{0.3cm}
{\em $^{\rm a}$Department of Physics,
P.O.Box 9, 00014 University of Helsinki, Finland\\}
\vspace{0.3cm}
{\em $^{\rm b}$Theory Division, CERN, CH-1211 Geneva 23,
Switzerland\\}
\vspace{0.3cm}
{\em $^{\rm c}$School of CPES, Univ. of Sussex, 
Brighton BN1 9QJ, UK\\}
\vspace{0.3cm}
{\em $^{\rm d}$NORDITA, Blegdamsvej 17,
DK-2100 Copenhagen \O, Denmark\\}
\vspace{0.3cm}
{\em $^{\rm e}$Department of Theoretical Physics, 
Lebedev Physical Institute,\\
Leninsky Prospect 53, 117924 Moscow, Russia}

\vspace{0.7cm}
{\bf Abstract}

\end{centering}

\vspace{0.3cm}\noindent
We study the phase diagram of the three-dimensional SU(3)+adjoint Higgs 
theory with lattice Monte Carlo simulations. A critical
line consisting of a first order line, a tricritical point and a
second order line, divides the phase diagram into two parts distinguished
by $\langle\tr A_0^3\rangle=0$ and $\not=0$. The location and the type of
the critical line are determined by measuring the condensates
$\langle\tr A_0^2\rangle$ and $\langle\tr A_0^3\rangle$,
and the masses of scalar and vector excitations. Although 
in principle there can be different types of broken phases, 
corresponding perturbatively to unbroken SU(2)$\times$U(1)
or U(1)$\times$U(1) symmetries, we find that dynamically 
only the broken phase with SU(2)$\times$U(1)-like properties
is realized. The relation of the phase diagram to 4d finite 
temperature QCD is discussed.
\vfill

\noindent
CERN-TH/98-350\\
NORDITA-98/66HE\\
November 1998

\vfill

\end{titlepage}

\section{Introduction}

The motivation for studying the three-dimensional (3d) SU(3) + 
adjoint Higgs gauge theory is twofold. First of all, this case 
is interesting since the 3d SU(3) + adjoint Higgs theory is an effective 
theory of finite temperature QCD in the weak coupling domain~\cite{dr}. 
The requirement of small couplings means that this effective theory 
is accurate only in the limit $T\gg\lambdamsbar$, not in the phase
transition region $T\approx T_c$ \cite{reisz}--\cite{dg}.
For pure gauge SU(3) theory, this is related to the fact 
that the phase transition at $T=T_c$ has to do with the breaking of the Z(3) 
symmetry. This symmetry is lost in the effective theory, but 
some traces of it remain, as will be discussed below.

The second interesting aspect of the 3d SU(3) theory is
that it already has several of the properties of the SU(5)
case, which is relevant for some GUTs at finite
temperatures~\cite{rajantiesu5}. These properties  
are not shared by the special SU(2) case, where the 
structure constants $d_{abc}$ vanish. Such properties
are the existence, in perturbation theory, of different
types of broken phases with the associated various gauge 
groups, and of the corresponding monopoles. However, 
the spectrum of various broken phases is of course 
richer in SU(5) than in SU(3), and  
the SU(5) action has also two non-trivial scalar self-couplings
in contrast to the SU(3) action, which only has one.
Nevertheless, we expect a similar qualitative behaviour.

Some conjectures concerning the phase diagram of the 3d SU(3) + 
adjoint Higgs theory were put forward in~\cite{bbk}. 
The purpose of this paper is to 
determine the phase diagram numerically. 

The actions of 3d SU($N\le 3$) + Higgs theories normally 
depend on two variables\footnote{As discussed below, 
in some cases there can in principle be more couplings than
just two.}:
the scalar mass $m_3^2$ and the scalar self-coupling $\lambda_3$. 
On the mean field level the system has two phases:
a ``symmetric'' phase at $m_3^2>0$ and a ``broken'' phase at $m_3^2<0$
at any $\lambda_3$. The problem now is to determine the phase
diagram in the full quantum theory. This question has 
previously been answered
with numerical lattice Monte Carlo simulations  
for a number of theories: SU(2) + fundamental representation Higgs
\cite{isthere}--\cite{endpoint}, SU(2)$\times$U(1) + 
fundamental Higgs \cite{su2u1}, and
SU(2) + adjoint Higgs \cite{hpst,su2adjoint}. 
The case of U(1) + fundamental scalar
Higgs (the Ginzburg-Landau theory) has also been
extensively  studied \cite{u1}. 

For small Higgs self-coupling, $\lambda_3\ll g_3^2$, the 
tree-level second order transition is, in all cases studied, 
radiatively changed into a first order transition. 
Its strength
decreases with increasing $\lambda_3$. The central and often very
difficult question is
what happens at larger $\lambda_3$: does the first order line terminate
(so that the two phases are analytically connected) or is there a
phase transition line across the whole phase diagram (so that there
is an order parameter vanishing in one of the phases). In this
respect,
the phase diagrams for the different symmetry groups differ 
in quite interesting ways from each other. A good illustration
of the difficulty of the problem is the fact that the nature
of the phase diagram in the superficially simplest case, the
Ginzburg-Landau theory, is not yet conclusively determined
in the large $\lambda_3$, extreme type II, regime.

For SU(3) + adjoint Higgs theory we shall see that the phase diagram
is divided into two parts by a phase transition line which contains
a first order line, a tricritical point and a second order line. In 
contrast, the SU(2) + adjoint Higgs theory is observed to have no
transition at large $\lambda_3$~\cite{hpst}.
The reason why it is possible to have a qualitatively 
different behaviour in SU(2) and SU(3) is  
that for all SU($N\ge3$) there is a gauge-invariant
local order parameter $\langle\tr A_0^3\rangle$ sensitive to the
breaking of the $A_0\leftrightarrow -A_0$ symmetry of the theory.

The plan of the paper is the following.
In \se\ref{sec:def} we define the theory in continuum 
and on the lattice. In \se\ref{sec:per} we present some
perturbative estimates for the different observables 
measured, and in \se\ref{sec:rel} we briefly review the 
relation to 4d finite temperature QCD. The simulation
results are in \se\ref{sec:sim}, and the conclusions 
in \se\ref{sec:con}.

\section{Definition of SU(3) + adjoint Higgs theory}
\la{sec:def}

The theory we study is defined by the following super-renormalizable
Lagrangian:
\ba
L[A_i^a,A_0^a] & = &  \fr14  F_{ij}^aF_{ij}^a
+ \tr [D_i,A_0][D_i,A_0] + 
m_3^2\tr A_0^2 +\lambda_A(\tr A_0^2)^2,
\la{leff}
\ea
in standard notation ($D_i=\partial_i+ig_3A_i$). The notation
$A_0=A_0^aT^a$ for the adjoint scalar is chosen with its origin
from dimensional reduction of QCD in mind. For SU(3) and SU(2) 
$\tr A_0^4=\fr12(\tr A_0^2)^2$ so that only one scalar self-coupling
appears. In principle, there could also be 
two additional super-renormalizable
couplings $h_3,h_5$ appearing as 
\be
\delta L = h_3 \tr A_0^3 + h_5 \tr A_0^2 \tr A_0^3. \la{h3h5}
\ee  
However, these terms
do not arise in the dimensional reduction of 
finite temperature QCD, and thus we assume that $h_3=h_5=0$.

In the absence of $h_3,h_5$, 
the theory in \eq\nr{leff} is symmetric under $A_0 \to -A_0$.
This symmetry was called ``R-parity'' in \cite{bbk}. In terms
of 4d physics, this symmetry is related to the usual
discrete transformations CT, P~\cite{ay,pviol}. 
However, it should be clearly stated
that the breaking of the $A_0 \to -A_0$ symmetry to 
be discussed below, does certainly not imply spontaneous
breaking of any of the discrete symmetries of 
finite temperature QCD, 
since the broken phases are
not physical from the point of view of QCD~\cite{su2adjoint}.

Due to super-renormalizability, 
of the couplings in \eq\nr{leff} only the scalar mass $m_3^2$ 
is scale dependent. In the $\msbar$ scheme
\be
m_3^2(\mu)={f_{2}\over16\pi^2}\ln{\Lambda\over\mu},\quad
f_{2}=20(3g_3^2-\lambda_A)\lambda_A,
\la{md}
\ee
where $\Lambda$ is a constant of dimension GeV.
The dynamics of the theory is thus determined by 
the dimensionless variables
\be
y={m_3^2(g_3^2)\over g_3^4},\quad
x={\lambda_A\over g_3^2},
\la{variables}
\ee
where $g_3^2$ has been used as a natural mass unit.

Instead of regulating the theory using the $\msbar$ scheme one can as
well use the lattice scheme with a lattice constant $a$. The two schemes
have to be matched so that they, in the limit $a\to0$ and for given
$g_3^2,y,x$, give the same physical results. Due to the super-renormalizability
this only requires tuning the bare mass term of the lattice action.
Introducing the gauge coupling constant $\beta_G$ by
\be
\beta_G={6\over ag_3^2},
\la{bg}
\ee
the lattice action $S=S[U_i(\bfx),A_0(\bfx)]$ becomes
\ba
S&=&\beta_G\sum_\bfx\sum_{i<j}(1-\fr13{\rm Re}\tr P_{ij}(\bfx)) \nn
&&-{12\over\beta_G} \sum_{\bfx,i}\tr A_0(\bfx)U_i(\bfx)A_0(\bfx+i)
U_i^\dagger(\bfx) \nn
&&+\sum_i\left\{\beta_2\ttwo(\bfx)+x{216\over\beta_G^3}[\ttwo(\bfx)]^2\right\},
\la{lattaction}
\ea
where $P_{ij}(\bfx)$ is the plaquette and
where the coefficient of the quadratic term is \cite{lc}
\ba
\beta_2&=&{36\over\beta_G}\left\{1+{6\over\beta_G^2}y-(6+10x)
{3.1759115\over
4\pi\beta_G}\right. \nn
&&\left.-{6\over16\pi^2\beta_G^2}[(60x-20x^2)(\ln\beta_G+0.08849)+
34.768x+36.130]\right\}. \la{2lm2}
\ea
Remarkably, the matching of lattice and continuum can be carried out
analytically even including terms of order $a$ \cite{moore}.
This implies that the $g_3^2,y,x$ in \eq\nr{lattaction} are modified
by corrections of order $1/\beta_G$. The additive
corrections for $y$ have not yet been
computed: the ${\cal O}(a)$ correction is, for dimensional
reasons, $\sim ag_3^6f(x,y)$ and a 3-loop computation would be needed.
For the two other parameters the improvements are
\ba
(g_3^2)_\rmi{improved} &=& \left(1+{1.994833\over\beta_G}\right)g_3^2,\nn
x_\rmi{improved}&=&x-{1\over\beta_G}(0.328432-0.835282x+1.167759x^2),
\la{ximp}
\ea
where $g_3^2,x$ are the parameters appearing in \eqs\nr{bg}--\nr{2lm2}.

The improved relations between the condensates of the scalar field
$A_0$ in the lattice action in \eq\nr{lattaction} and in 
the continuum are\footnote{We
thank Guy Moore for providing us with the improved expression of $\langle
A_0^3\rangle $.}
\ba
{\langle\tr A_{0}^2\rangle_\rmi{cont}\over g_3^2} &=& 
\bigg(\langle\tr A_0^2\rangle
-{3.1759115 \beta_G\over6\pi}\bigg)
\bigg(1-\fr{3.409891-0.729850x}{\beta_G}\bigg)\nn
&&-{3\over2\pi^2}(\ln\beta_G+0.6678), \la{tra02} \\
{\langle\tr A_{0}^3\rangle_\rmi{cont}\over g_3^3} &=& 
\langle\tr A_0^3\rangle
\bigg( 1 - \fr{5.114364-0.437910x}{\beta_G} \bigg).
\la{tra03}
\ea
In \eq\nr{tra02} an ${\cal O}(ag_3^2)$ additive 3-loop correction is
still not known. The $1/\beta_G$-terms are numerically quite large
at the value $\beta_G=12$ we have used in practice (up to $\sim 40\%$), 
and implementing them brings the lattice results
significantly closer to the perturbative results
(deep in the perturbative regime). 

The $\msbar$-scheme regularized operator
$\langle\tr A_{0}^2\rangle_\rmi{cont}$ is, 
in fact, scale dependent~\cite{framework}, and has been defined at the scale
$\mu=g_3^2$ in \eq\nr{tra02}. 
The scale dependence arises at the 2-loop level and
is, for general $N$,  
\be
\trasq(\mu')=
\trasq(\mu)+{g_3^2(N^2-1)N\over16\pi^2}\ln{\mu'\over\mu}.
\la{expval}
\ee
In terms of the effective potential 
(note that $\langle\tr A_{0}^2\rangle_\rmi{cont}=dV(\mbox{min})/dm_3^2$),
the scale dependence arises from the graph (solid line = $A_0$,
wavy line = $A_i$)

\vspace*{0.2cm}

\be
\parbox[c]{1.5cm}{
\begin{picture}(30,30)(0,0)
\SetWidth{1.0}
\SetScale{1.2}
\Photon(0,15)(30,15){1.5}{4}
\CArc(15,15)(15,0,360)
\end{picture}}
\propto
{g_3^2(N^2-1)N\over16\pi^2}m_3^2
\biggl(\ln{\mu\over2m_3}+\fr34\biggr). 
\la{SSV}
\ee

\vspace*{0.2cm}

\noindent
It is amusing to note that this term is precisely the
${\cal O}(g^4\ln(g))$ Toimela term \cite{toimela,bn} in 
the free energy of the QCD plasma. 

The gauge invariant operators of lowest dimensionalities
in the action defined by \eq\nr{lattaction} are as follows:
\bi
\item Dim=1: $\ttwo$,
\item Dim=3/2: $\tthree$,
\item Dim=2: $(\ttwo)^2$, $\epsilon_{ijk}\tr A_0 F_{jk}$,
\item Dim=5/2: $\epsilon_{ijk}\tr A_0^2F_{jk}$,
\item Dim=3: $\tr F_{ij}F_{kl}$, $\tr [D_i,A_0]\,[D_j,A_0]$ (with various
spin projections).
\ei
These operators can be used for mass measurements
in the different quantum number channels. 

\section{Perturbation theory}
\la{sec:per}

Because of confinement, perturbation theory is not well convergent
in the symmetric phase of the theory and is thus, in general, of limited
usefulness in the study of the phase structure.
Nevertheless, perturbation theory does work in the limit of 
small $x=\lambda_A/g_3^2$ when the transition becomes very
strong, and it is worthwhile to go through its predictions there.

Let us first fix a gauge and parametrise 
a constant diagonal SU(3) background field as follows:
\be
g_3^{-1} \langle A_0\rangle=B=B_3T^3+B_8T^8=
\left( \begin{array}{ccc}
q+p & 0 & 0 \\
0 & q-p & 0 \\
0 & 0 & -2q
\end{array} \right),
\la{background}
\ee
with
\be
\ttwob = 6q^2+2p^2,\quad \tthreeb=6q(p^2-q^2),\quad \tfourb=\fr12(\ttwob)^2.
\la{traces}
\ee
(Note that $\tr B^2 = g_3^{-2} \tr \langle A_0\rangle^2
\neq g_3^{-2} \langle \tr A_0^2 \rangle.$)
If $p=0$ or $p=\pm 3q$, then 
the SU(3) symmetry is broken to SU(2)$\times$U(1), otherwise
it is broken to U(1)$\times$U(1). These are perturbative statements;
the only symmetry that can be broken in the full quantum theory
is $A_0 \leftrightarrow -A_0$, which is signalled by $\langle\tthree\rangle$.

\subsection{The $x\to0$ limit}

Let us 
compute the 1-loop effective potential $V_1(x,y;q,p)$ in the background
in \eq\nr{background}. One finds, in the Landau gauge,
\ba
&&g_3^{-6}V(x,y;q,p)=y(6q^2+2p^2)+x(6q^2+2p^2)^2\nn
&&-{1\over3\pi}(|3q+p|^3+|3q-p|^3+8|p|^3)\nn
&&-{1\over12\pi}\left\{[y+6x(6q^2+2p^2)]^{3/2}+7[y+2x(6q^2+2p^2)]^{3/2}
\right\}.
\la{v1loop}
\ea
The scalar loop terms (last line in \eq\nr{v1loop}) become negligible when
$x\to0$ (apart from a constant)
and we shall neglect them to begin with. Then, for $p=0$,
\be
g_3^{-6}V(x,y;q,0)=36xq^2\left[\left(|q|-{1\over4\pi x}\right)^2+
{1\over6x}\left(y-{3\over8\pi^2x}\right)\right].
\la{v(q0)}
\ee
Thus, for
\be
y=y_c^\rmi{1loop}(x)={3\over8\pi^2x}
\la{crline}
\ee
the system has two coexisting states (a first order transition) with
\be
q_\rmi{symmetric}=0,\qquad q_\rmi{broken}={1\over4\pi x}.
\ee
One stable and one metastable state exists for
\be
y_-(x)=0 < y < y_+(x)={9\over8}y_c(x).
\ee

These results for $p=0$ can now
be extended to the whole $(q,p)$-plane. Indeed, 
\be
V_1(x,y;q>0,p=3q)=V_1(x,y;2q,0),\quad V_1(x,y;q<0,p=-3q)=V(x,y;2q,0).
\ee
This implies that, at $y=y_c(x)$, the system has the following
degenerate potential minima (Fig.\ref{minima}):
\bi
\item[(0)] a symmetric minimum at $q=0$, $p=0$,
\item[(1)] a broken minimum at $q=1/(8\pi x),\,\,p=3/(8\pi x)$,
\item[(2)] a broken minimum at $q=1/(4\pi x)$, $p=0$.
\ei
Moreover, in the parametrisation in \eq\nr{background} one has the freedom
of permuting the diagonal elements and the potential has to be
invariant under the transformations of $q,p$ corresponding to these
permutations. One sees that the fundamental region, which determines the
potential over the whole plane, can be chosen to be that
bounded by the two lines $p=0,\,p=3q$. Thus, for
each broken minimum there are two more minima
corresponding to cyclic permutations of the diagonal elements.  
All these minima correspond to breaking to SU(2)$\times$U(1);
there are no local minima corresponding to breaking to U(1)$\times$U(1)
(a U(1)$\times$U(1) minimum would require that all the 
vector masses cubed appearing on the 2nd line in \eq\nr{v1loop}
are non-vanishing, but this is not the case in any of the 
minima considered).

\begin{figure}[t]


\centerline{ 
\epsfxsize=8cm\epsfbox{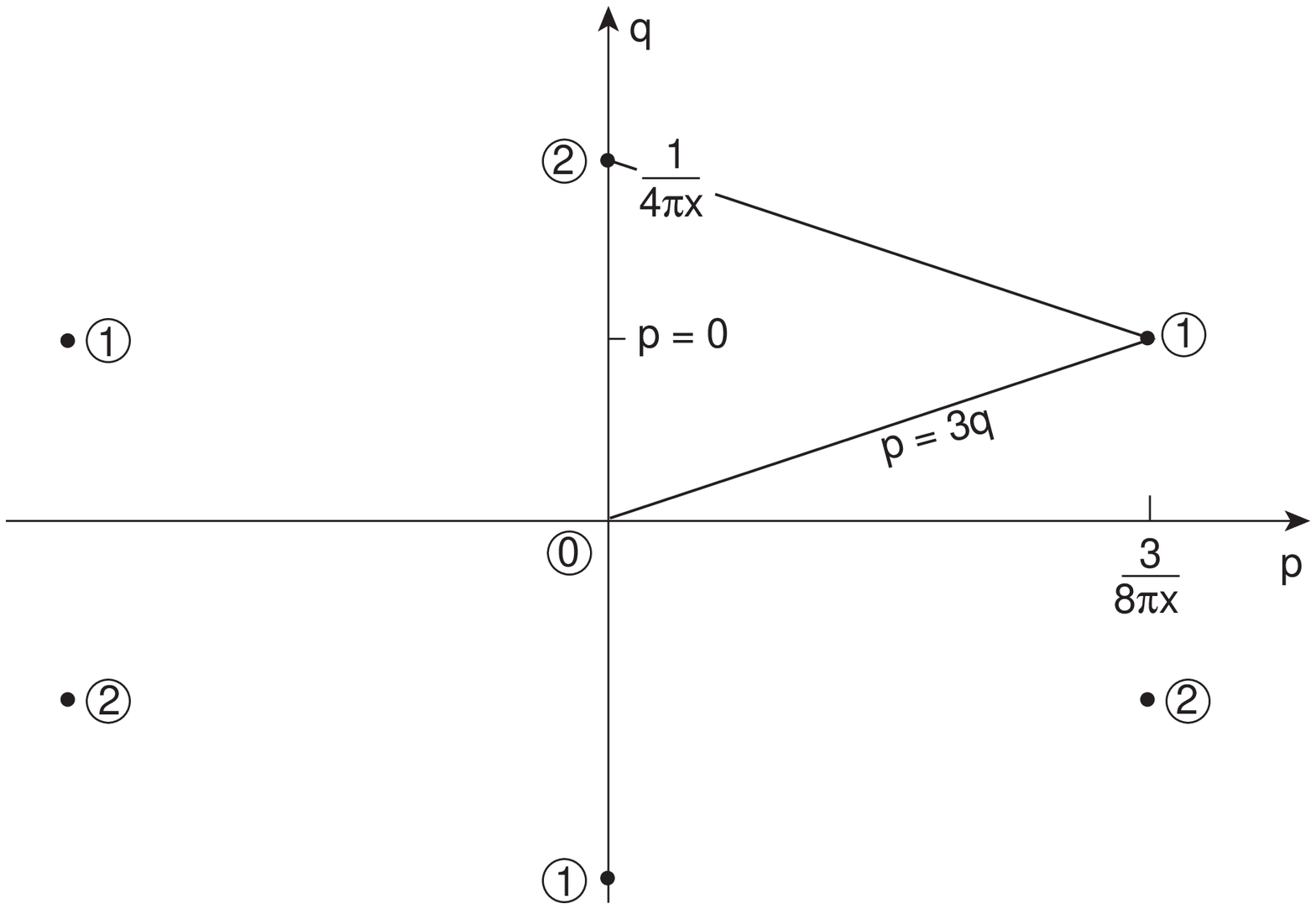} ~~~
\epsfxsize=7.5cm\epsfbox{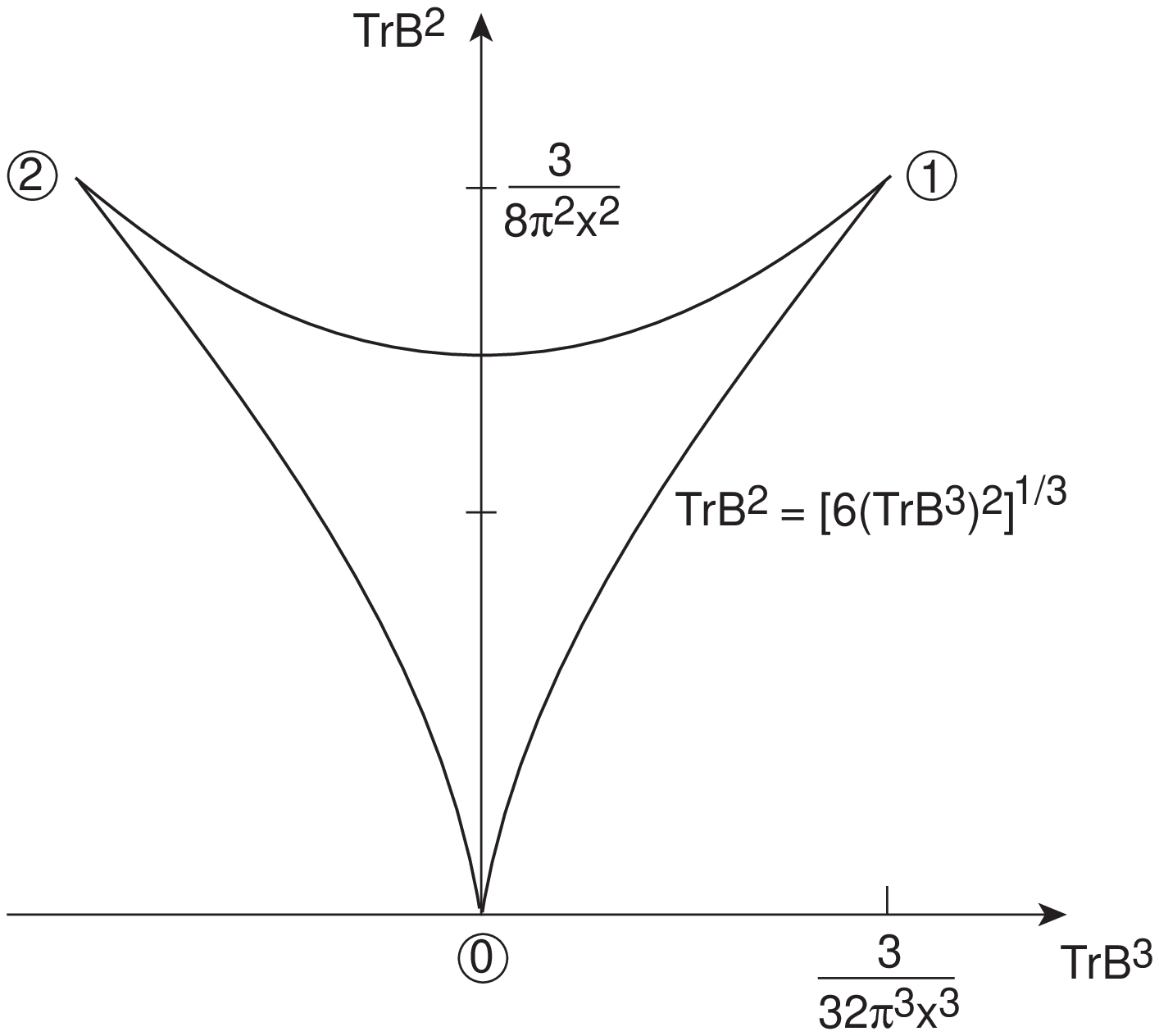}}

\caption[a]{The minima of the 1-vectorloop potential on the critical
line $y=y_c(x)$ (\eq\nr{crline}). {\em Left:} The minima
in the $(q,p)$-plane. Those with the same number correspond to permutations
of the eigenvalues of $B$ in \eq\nr{background}. The fundamental region is
bounded by the two lines $p=0,\,p=3q$.
{\em Right:} The minima in the $(\ttwob,\tthreeb)$-plane. The boundary curve
is the map of the triangle in the left panel.
}
\la{minima}
\end{figure}

To get rid of the unphysical extra symmetries related to 
the permutation of the diagonal elements of $B$ in \eq\nr{background},
it is useful to write the 1-loop potential in terms of $\ttwob$ and
$\tthreeb$. Effectively, one inverts the cubic \eqs\nr{traces} 
and inserts into \eq\nr{v1loop}. The result is 
\ba
&&V_1(x,y;\ttwob,\tthreeb) = y \ttwob + x (\ttwob)^2 \nn
&& - {1\over3\pi} \left\{
\ttwob{\rm Re}\left[\left(\fr92\tthreeb+\fr32\sqrt{3}i  
\sqrt{\fr12(\ttwob)^3 - 3(\tthreeb)^2}\right)^{1/3} (3-i\sqrt{3})\right]
\right.\nn
&&-\left.\sqrt{\fr12 (\ttwob)^3 - 3(\tthreeb)^2}\right\} 
\Theta\Bigl[\fr12 (\ttwob)^3 - 3(\tthreeb)^2 \Bigr].
\la{t2t3boundary}
\ea
This form, containing the solution of a cubic equation,
is not very transparent, but it actually is quite simple,
as shown in Fig.\ref{minima}. Now only the genuinely different minima,
one symmetric and two broken ones, appear. 

Summarising, for $x\to0$ perturbation theory predicts that the
system has the following three phases:
\bi
\item
one symmetric phase with $\ttwob=\tthreeb=0$
for $y>y_c^\rmi{1loop}(x)$,
\item two broken phases for $y<y_c^\rmi{1loop}(x)$, distinguished
by the sign of $\tthreeb$.
\ei
In the broken phase at the critical point $y=y_c$, 
\ba
\ttwob &=& {3\over8\pi^2x^2}\approx3.80 \left({0.1\over x}\right)^2,\nn
\tthreeb &=&\pm{3\over32\pi^3x^3}\approx \pm 3.02
\left({0.1\over x}\right)^3.
\la{brokenminima}
\ea

\subsection{Condensates}

Using the effective potential in \eq\nr{v1loop}, 
one can calculate the 1-loop
perturbative approximation for $\trasq$,
\be
g_3^{-2}\trasq=g_3^{-6}{\partial V\over\partial y}
=6q^2-\frac{1}{8\pi}
\left[(y+36xq^2)^{1/2}+7(y+12xq^2)^{1/2}\right],\la{pert2}
\ee
where $q$ is obtained by minimizing the potential 
in \eq\nr{v1loop} with $p=0$
and is zero in the symmetric phase. A similar calculation yields
\be
g_3^{-3}\langle\tthree\rangle
=\pm\left\{
6q^3-\frac{3q}{8\pi}
\left[(y+36xq^2)^{1/2}-(y+12xq^2)^{1/2}\right]\right\}. \la{pert3}
\ee
At the limit $x\to 0$ considered
in \eq\nr{brokenminima}, $g_3^{-2}\trasq= \tr B^2 -(y^{1/2}/\pi)$, 
$g_3^{-3}\langle\tthree\rangle = \tr B^3$. Note that the 
perturbative $\msbar$ value of $\trasq$ is 
negative in the symmetric phase.

\subsection{What happens at larger $x$?}
When $x$ increases, the scalar contributions
become more important, 
perturbation theory becomes less accurate, the
three well separated phases in Fig.\ref{minima} approach each other,
and the transition gets weaker. In the case of the SU(2) + adjoint Higgs
theory this leads to an endpoint of the first order line~\cite{hpst}: 
there is no local order parameter, and the phases are believed to
be analytically connected. In the present case, the transition 
also becomes weaker with
increasing $x$.  However, now there is a gauge
invariant local order parameter, $\langle\tthree\rangle$, which can
signal the breaking of the Z(2) symmetry $A_0\to -A_0$ of the
theory. Thus we expect that the $(x,y)$-plane phase diagram can be
disconnected by a critical line containing a first order transition at
$x<x_c$, a {\em tricritical point\,} at $x=x_c$, and a second order
line at $x>x_c$.

However, as is standard in the case of tricritical transitions, in order
to see the full phase diagram one has to use three couplings/fields.  In
our case the couplings are $x$, $y$ and $h$, an external field with a
coupling $h \tr A_0^3$ (see \eq\nr{h3h5}).  
The schematic 3-dimensional phase diagram is
shown in the left panel of \fig\ref{fig:tricritical}.  We note that 
in this full coupling space the
three phases are still analytically 
connected through the large $x$ region.

\begin{figure}[t]


\centerline{ 
\epsfxsize=7cm\epsfbox{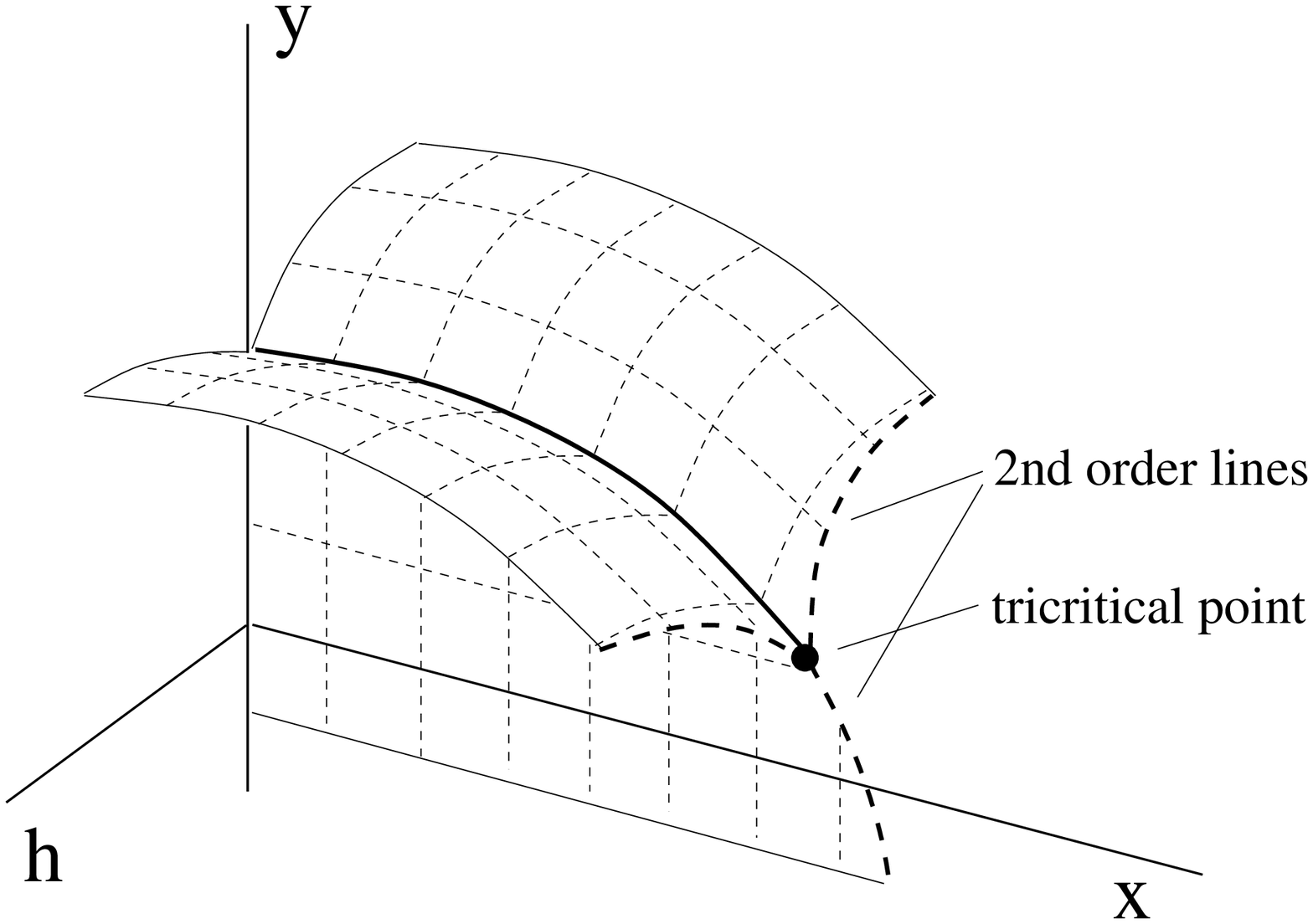}~~~~\epsfxsize=7cm\epsfbox{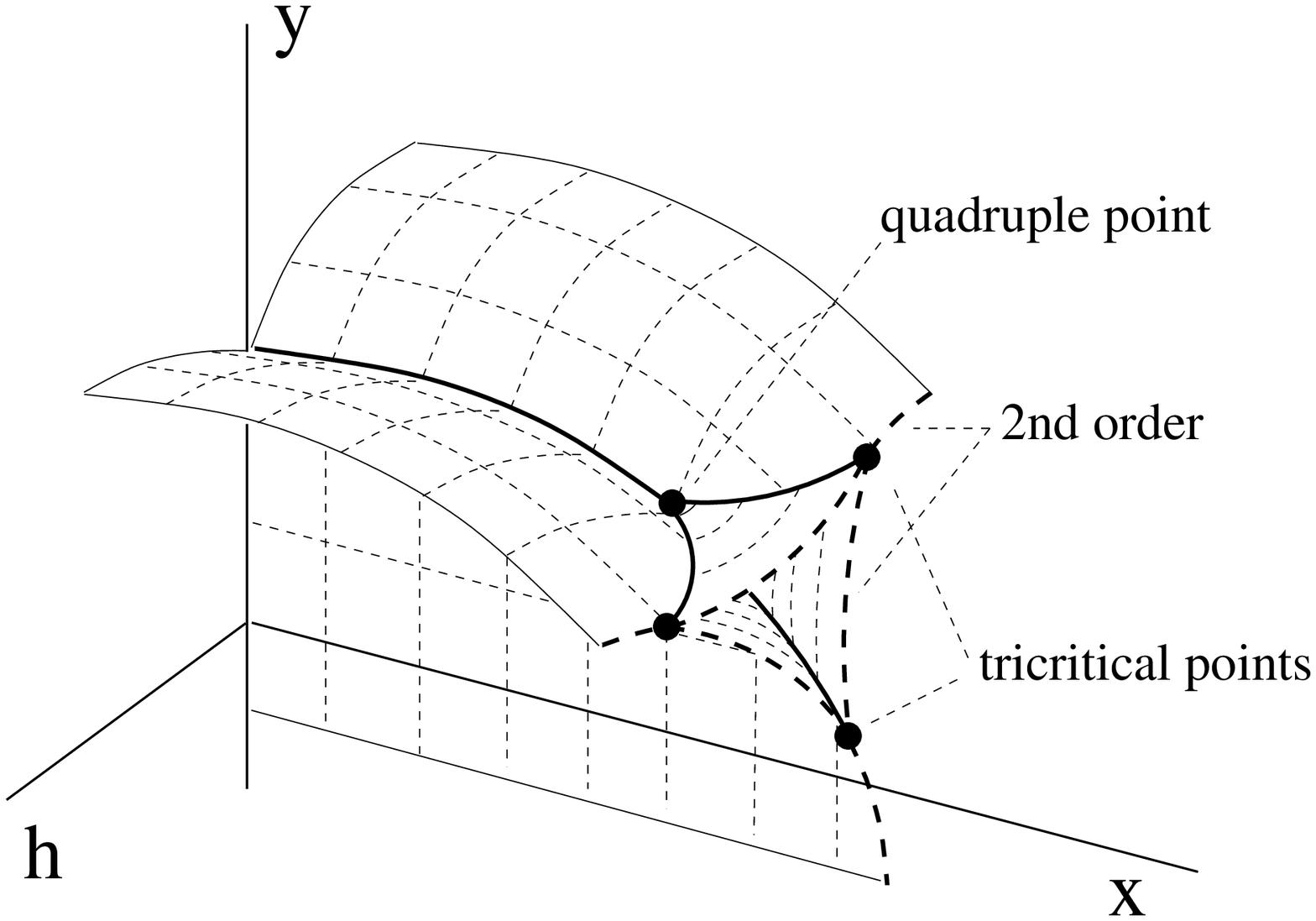}}
\caption[a]{Schematic phase diagrams of systems with a tricritical
point (left) and a quadruple point (right), shown in a 3-dimensional
coupling space $(x,y,h)$, where $h$ is an external field which couples
to $\tr A_0^3$.  The surfaces are 1st order phase transitions, and
dashed lines 2nd order (Ising) transitions.  The 1st order surfaces
join along lines of triple points, where three phases are separated by 1st
order phase transitions.  At the quadruple point (right) four phases are
separated by 1st order transitions; this is an intersection point of four
lines of triple points and six surfaces of 1st order transitions.}
\la{fig:tricritical}
\end{figure}

In principle, more complicated phase structures are also possible.  At
large enough $x$ a new phase with $\langle \tr A_0^3 \rangle = 0$,
$\langle \tr A_0^2 \rangle > 0$ may appear (compare with the symmetry
breaking patterns shown in \eq\nr{traces}!). There is no local
order parameter separating this phase form the symmetric one, but
perturbatively it corresponds 
to a phase with the gauge group U(1)$\times$U(1).
If this occurs, there will be a quadruple point at some
value of $x$, where four phases separated
by 1st order transitions can exist (see the right panel of
\fig\ref{fig:tricritical}).  The phase structure suggested in
\cite{bbk} belongs to this class, even though the authors
discussed the phases only in the $(x,y)$-plane.

This kind of phase structure with a quadruple point is known to exist
in some theories, for example, in the 3d 3-state Potts model
\cite{3d3spotts} and, indeed, in the finite temperature pure SU(3)
gauge theory in 3+1 dimensions.  These models have an exact Z(3)
symmetry, and they have 3 degenerate broken phases and 1 unbroken
phase, which can all exist 
at the quadruple point.  For the case of
SU(3) gauge theory, the 3d phase diagram is spanned by $1/T$ and
external fields coupling to the real and imaginary parts of the
Polyakov line.  Indeed, the phase space of the Polyakov line in the
complex plane does resemble the phase structure of the SU(3) + adjoint
Higgs theory, shown in \fig\ref{minima}.  If one considers the SU(3) +
adjoint Higgs theory as a dimensionally reduced version of the 4d
gauge theory, it is appealing to think that the phase structures of
the two theories could be similar.  However, one has to bear in mind
that in the 3d adjoint Higgs theory the 3-fold symmetry of the
original 4d theory is strongly broken, as will be discussed in the
next section.  Even then it is not impossible that the Z(3) symmetry
is dynamically generated in the close proximity of the would-be
tricritical point (this can be compared with the dynamically generated
Ising symmetry in the SU(2) gauge + fundamental Higgs theory
\cite{endpoint}).  If this truly happens, the would-be tricritical
point could transform into a quadruple point.


The question about the nature of the phase diagram is settled
numerically below, and it will turn out that the phase diagram is the
standard tricritical one, as shown in the left part of
\fig\ref{fig:tricritical}.

\section{Relation to 4d and to Z($N$) symmetry}
\la{sec:rel}

The relation between $x,y$ and 4d physics ($T,\lambdamsbar, N,N_f$)
is worked out explicitly to leading + next-to-leading order in
\cite{su2adjoint,debye}. For $N=3,N_f=0$ the answer is
\ba
x&=&{3\over8\pi^2}g^2(4\pi e^{-\gamma_E-3/11}T)\nn
&=&{3\over11}\,\,{1\over\ln(5.371T/\lambdamsbar)}, \\
y&\equiv& y_\rmi{dr}(x)={3\over8\pi^2x}+{9\over16\pi^2}+{\cal{O}}(x).
\la{dr}
\ea

As discussed, 
the 4d finite $T$ theory without matter in the fundamental 
representation possesses an extra
symmetry, the Z($N$) symmetry, for which the thermal Wilson line
\be
L(\bfx)={1\over N}\tr P\exp\left[ig\int_0^\beta d\tau A_0(\tau,\bfx)
\right]
\la{wilson}
\ee
is an order parameter: if $z_k=\exp(i2\pi k/N)$  one can define a
transformation of the fields so that the action is invariant but 
$L\to z_kL$. Thus $\langle L(\bfx)\rangle$ acts as an order
parameter.

This symmetry, however, is lost in the reduction process. The reason
is simply that in the reduction process mass scales $\sim\pi T$ are
integrated over. However, under a Z($N$) transformation a field configuration
with small fields is transformed to one with $A_0\sim \pi T/g$ and
the reduction process cannot accurately represent such large scales.

Even though the exact symmetry is lost, 
part of it is restored by radiative effects in the
effective theory. The leading order term in \eq\nr{dr} coincides
with the leading order critical line \eq\nr{crline}. This can be seen
as a remnant of the Z($N$) symmetry, since in this approximation,
the effect of the Z($N$) transformations of the 4d theory is to move the
system from one of the three degenerate minima of 
\fig\ref{minima} to another. 
This degeneracy is lost when higher-order corrections
are taken into account, and the true transition line
of the 3d theory does not agree
with the dimensional reduction line.
Moreover, in the effective 3d theory, only the vertices
marked by 1 and 2 are related by the symmetry of the theory;
in hot QCD (and in the 3d 3-state Potts model) all the vertices
are equivalent. It is thus also natural that in the effective
theory the middle point of the triangular region in \fig\ref{minima}
plays no special role, in contrast to the situation in hot QCD,
where the middle point corresponds to the confined phase.

\section{Simulation results}
\la{sec:sim}

The primary aim of the simulations is to resolve the nature of the
phase diagram and find the $(x,y)$-plane critical curve $y=y_c(x)$.
We also measure the screening masses (inverse screening lengths)
on both sides of the transition line at various values of $x$.

\begin{table}[bt]
\newcommand{\tube}[2]{~~$#1^2\times #2$}
\newcommand{\cube}[1]{~~$#1^3$}
\newcommand{\mc}{_m}
\center
\begin{tabular}{cl} 
\hline
 $x$  &    ~~Volumes  \\
\hline
0.10 &  \cube{12} \\
0.15 &  \cube{16} \\
0.20 &  \cube{16},\cube{32} \\
0.25 &  \cube{16},\cube{24},\cube{32},\cube{48} \\
0.30 &  \cube{16},\cube{32} \\
\hline
\end{tabular}
\caption[0]{The lattice sizes used at $\beta_G=12$.  For $0.15\le x\le
0.30$ several values of $y$ were used for each $x$, both above and
below the transition.  For $x\le0.20$, multicanonical simulations were
used at $y=y_c(x)$.}\label{tab:lattices}
\end{table}

The simulations were performed using $\beta_G = 6/(g_3^2 a) = 12$,
with $x$ in the interval $0.10\le x\le 0.30$.  The use of only one
lattice spacing (one $\beta_G$) precludes the extrapolation of the results
to the continuum limit.  However, we expect the finite lattice spacing
effects to be small enough for the purpose of mapping the phase
diagram.  This is supported by our experiences from the measurement
of the phase diagram of the SU(2) + adjoint Higgs model \cite{su2adjoint}.

The lattice volumes and the values of $x$ used in this study are
listed in Table~\ref{tab:lattices}.  For each $x$ and lattice size,
several runs at various values of $y$ were performed.  The total
number of runs was 72, with approximately 2.1 node-years of cpu-time on
a Cray T3E\@.  This corresponds to $\sim 2\times10^{16}$ floating point
operations at 130 Mflops/node.

\subsection{Local order parameters}
\la{sec:loc}

For $x\le 0.2$, the transition point $y=y_c(x)$ was determined with
multicanonical simulations using only lattices of size $12^3$ and
$16^3$.  The transition here is so strongly of the first order -- that is,
the latent heat and the surface tension are so large -- that the system
tunnels from one phase to another too infrequently, even when
multicanonical simulations are used.  The value of $y_c(x)$ was
determined to be the value of $y$ where the probability weight of each
of the 3 phases is equal.

\begin{figure}[t]
\centerline{\hspace{0.2cm} $x=0.15$ 
  \hfill\hfill\hfill $x=0.20$ \hfill\hfill\hfill}
\vspace*{-2mm}
\centerline{\epsfxsize=10cm\epsfbox{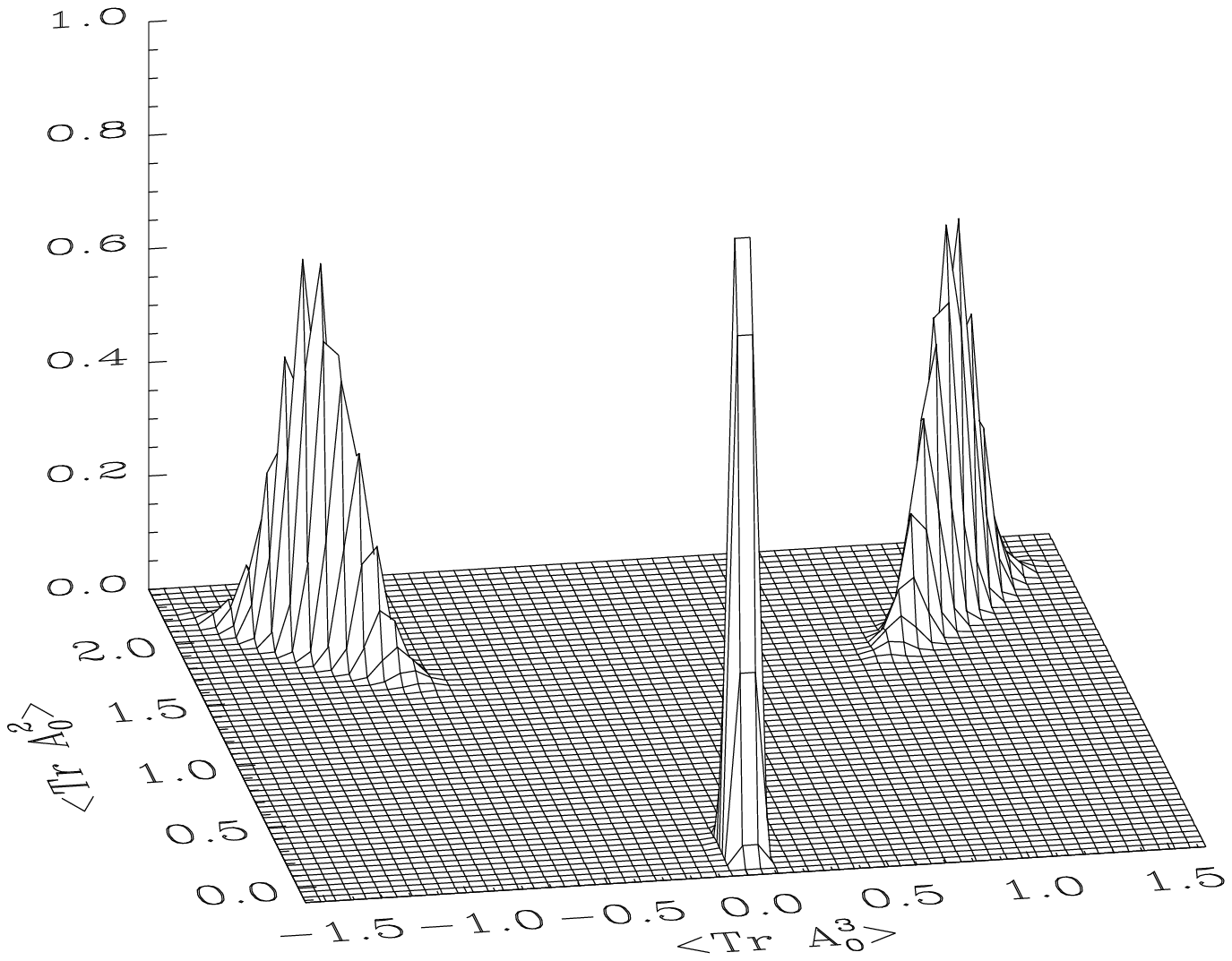}
\hspace{-2cm}\epsfxsize=10cm\epsfbox{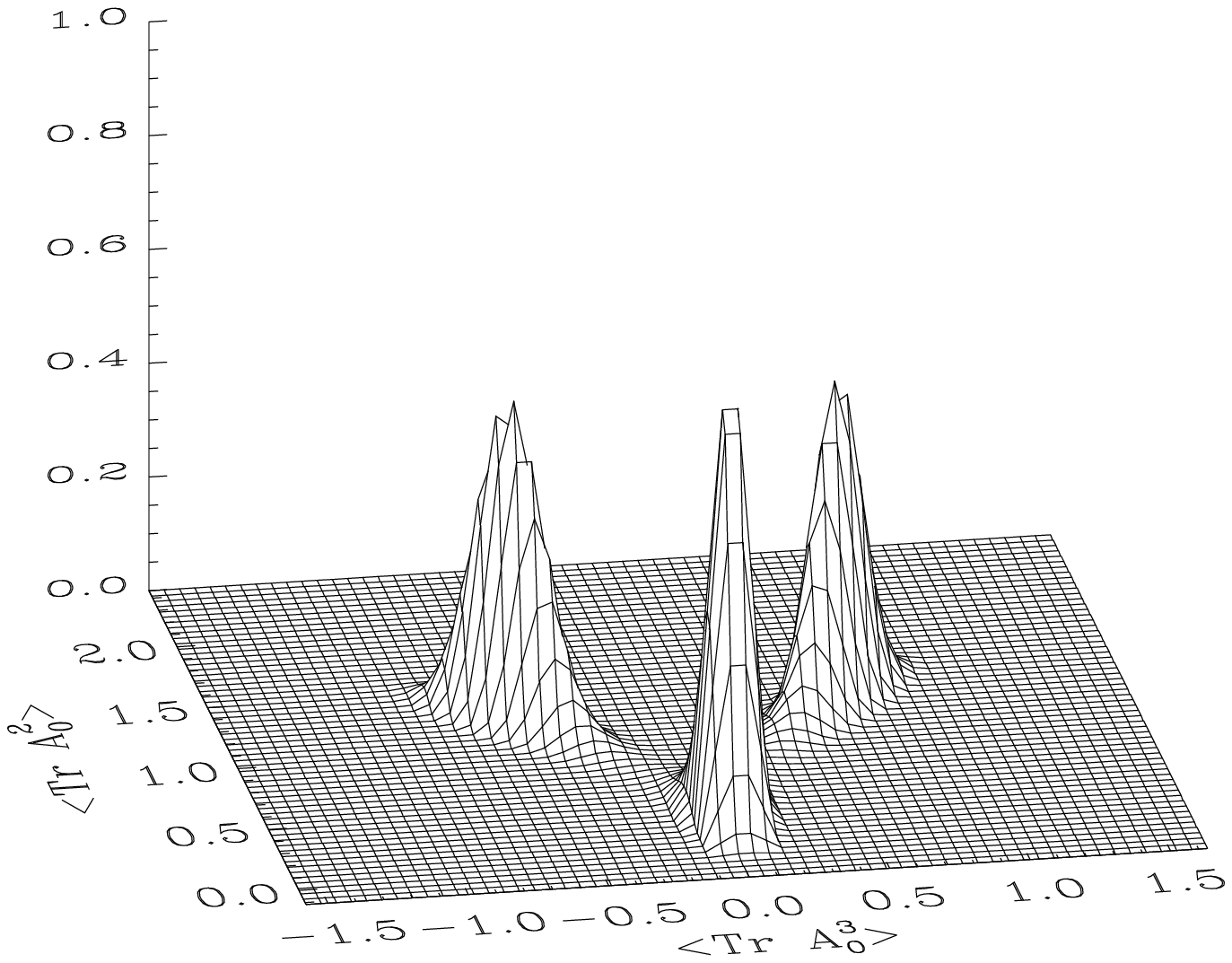}}
\centerline{ $x=0.25$ \hspace{5cm}}
\vspace{-2mm}
\centerline{\epsfxsize=10cm\epsfbox{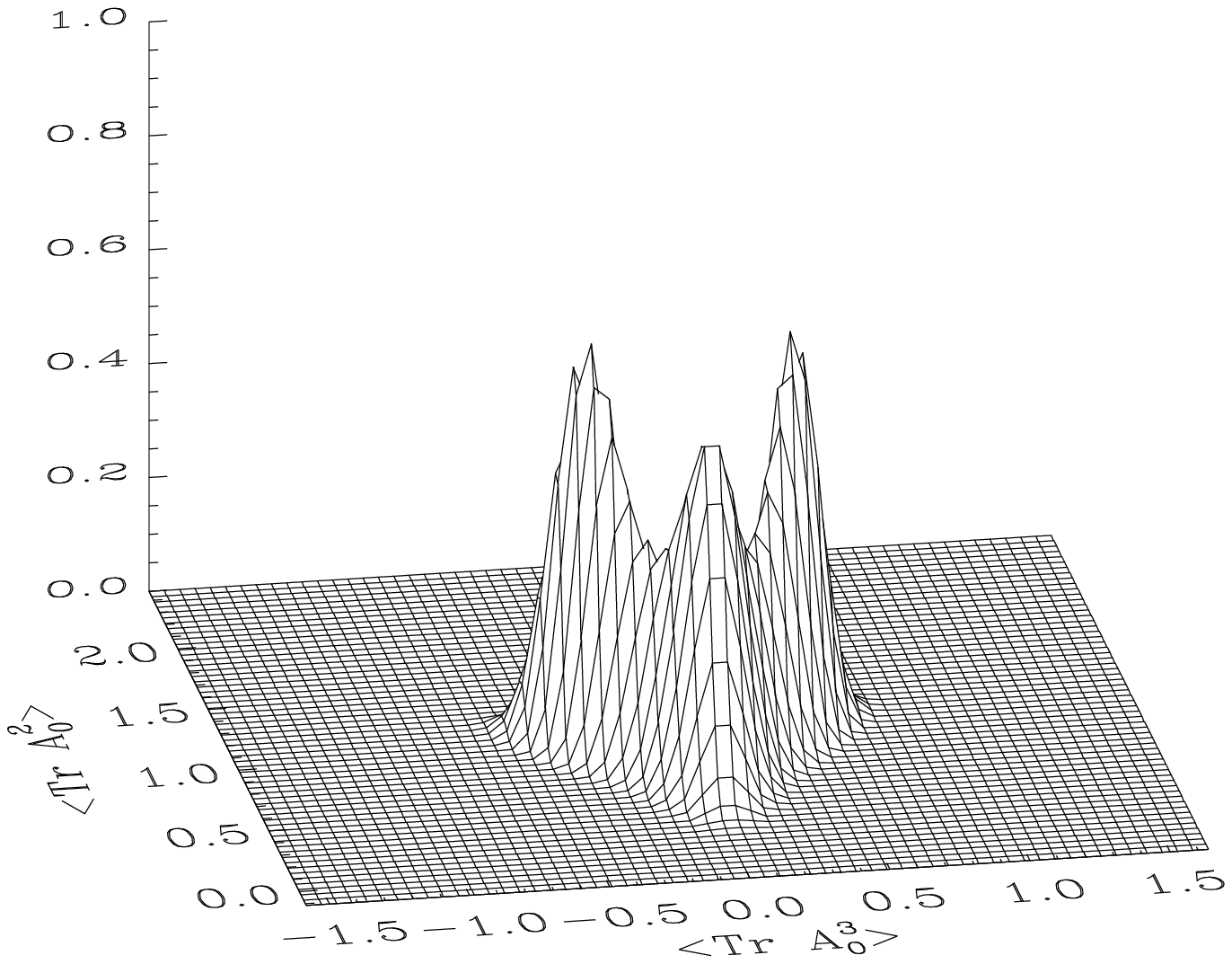}}
\caption[a]{The probability distributions in the 
$(\tthree,\ttwo)$-plane at $y=y_c(x)$, $x=0.15$, $0.20$, $0.25$.  
The data is from $16^3$, $\beta_G=12$ simulations;
the simulations at $x=0.15$
and $x=0.20$ are multicanonical.}
\la{distributions}
\end{figure}

The strength of the transition is illustrated in
\fig\ref{distributions}, where we show the probability
distribution in the $(\tthree,\ttwo)$-plane on the transition line for
$x=0.15$...0.25\@.  For small $x$, the three peaks are very strongly
separated, but when $x$ approaches 0.25, the peaks join.  It
should be noted that in each case, the two $\ttwo >0$ peaks are always
connected by a ``tunnelling channel'' to the $\ttwo \approx 0$ phase.
When $x=0.15$, the relative probability density in these channels is
suppressed by a factor $\sim 10^{-9}$ when compared with the peaks,
making them utterly invisible in \fig\ref{distributions},
plotted with a linear scale.  At $x=0.2$,
this suppression is ``only'' a factor of $\sim 0.01$, and at $x=0.25$,
there is no significant suppression any more.  Note that
the magnitude of the suppression is not universal and it depends very
strongly on the volume of the system.  Indeed, in large enough volumes
the magnitude of the suppression can be related to the interface
tension $\sigma$ between the phases.  However, the volumes used here are
too small for a reliable determination of $\sigma$.

It should be noted that 
there is no ``tunnelling channel'' directly connecting the two $\ttwo >0$
peaks even near the tricritical point.  Thus, the tunnelling from one
of these peaks to the other always proceeds through the region of the
$\ttwo \approx 0$ peak.  This indicates that there is {\em no\,}
dynamical generation of the Z(3) symmetry, and the behaviour
is the standard tricritical one.

In fact, we did not observe any sign of the quadruple point and the 
associated new phase in any point of the $(x,y)$-plane
(since $\langle\ttwo\rangle,\langle\tthree\rangle$ are
always simultaneously small or large, see below).  
This indicates that the phase diagram shown in the right panel of
\fig\ref{fig:tricritical} is not relevant here.

\begin{figure}[tb]
\centerline{ 
\epsfxsize=6.5cm\epsfbox{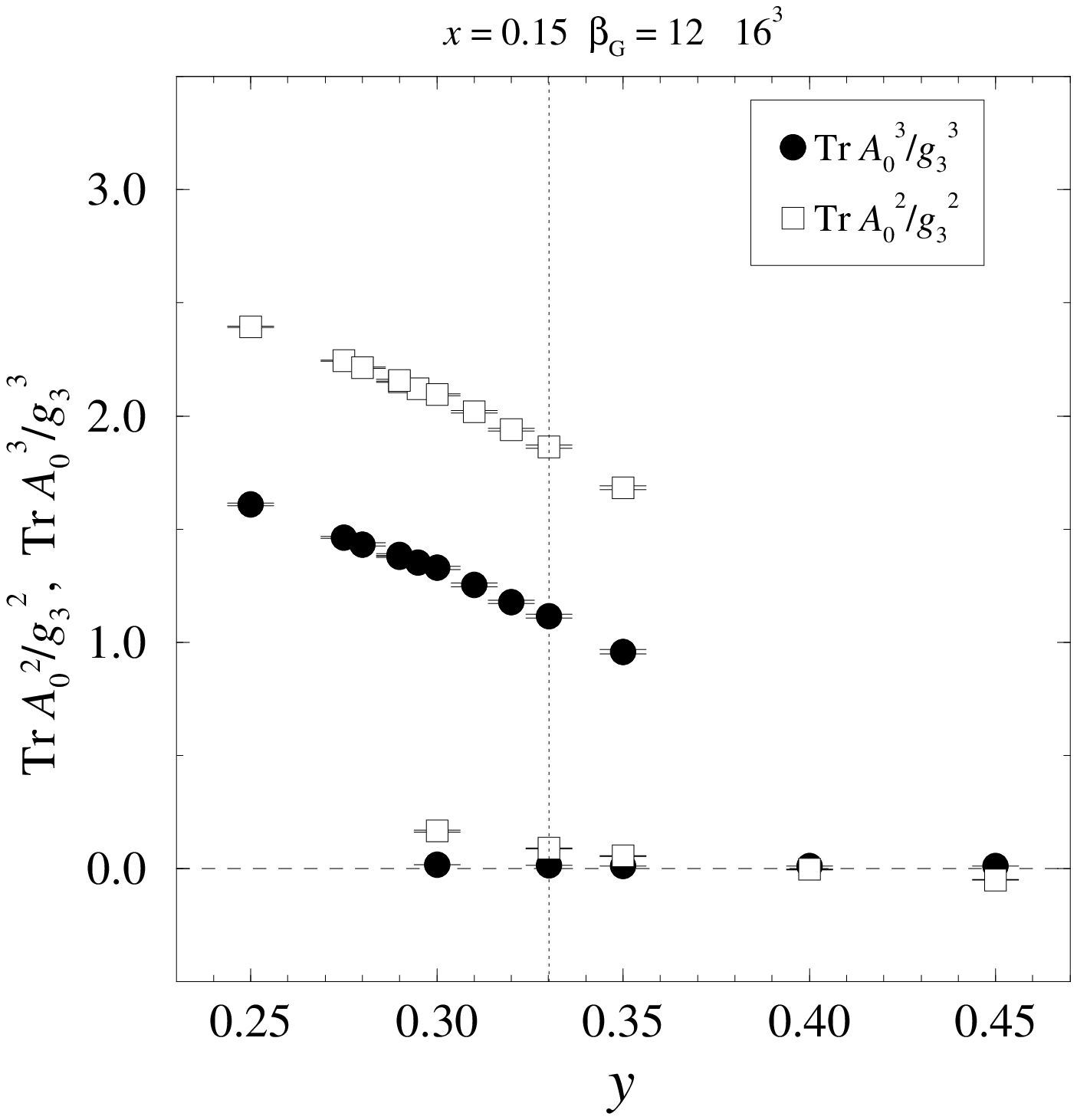} ~~~
\epsfxsize=6.8cm\epsfbox{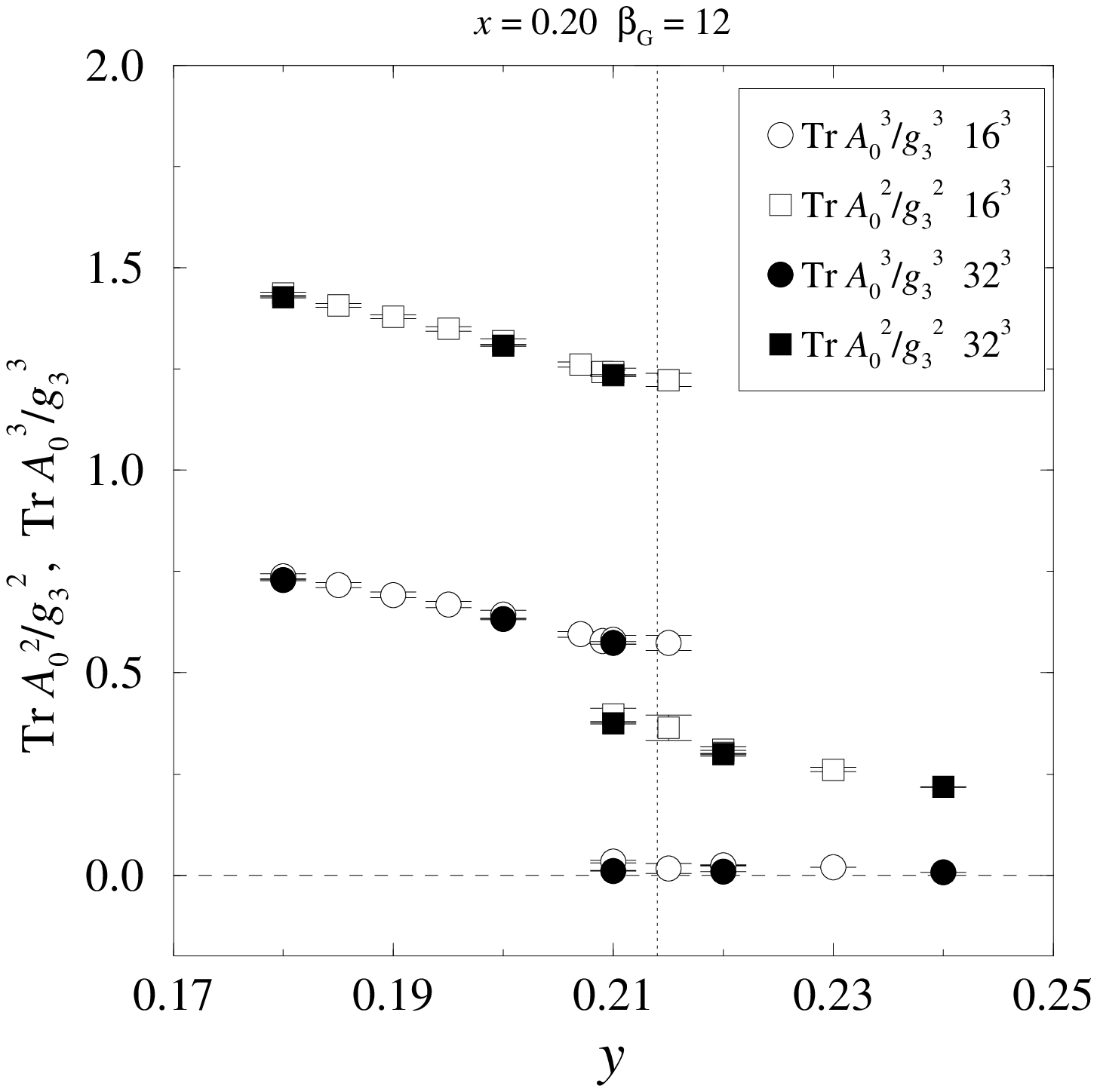}}
\centerline{
\epsfxsize=6.5cm\epsfbox{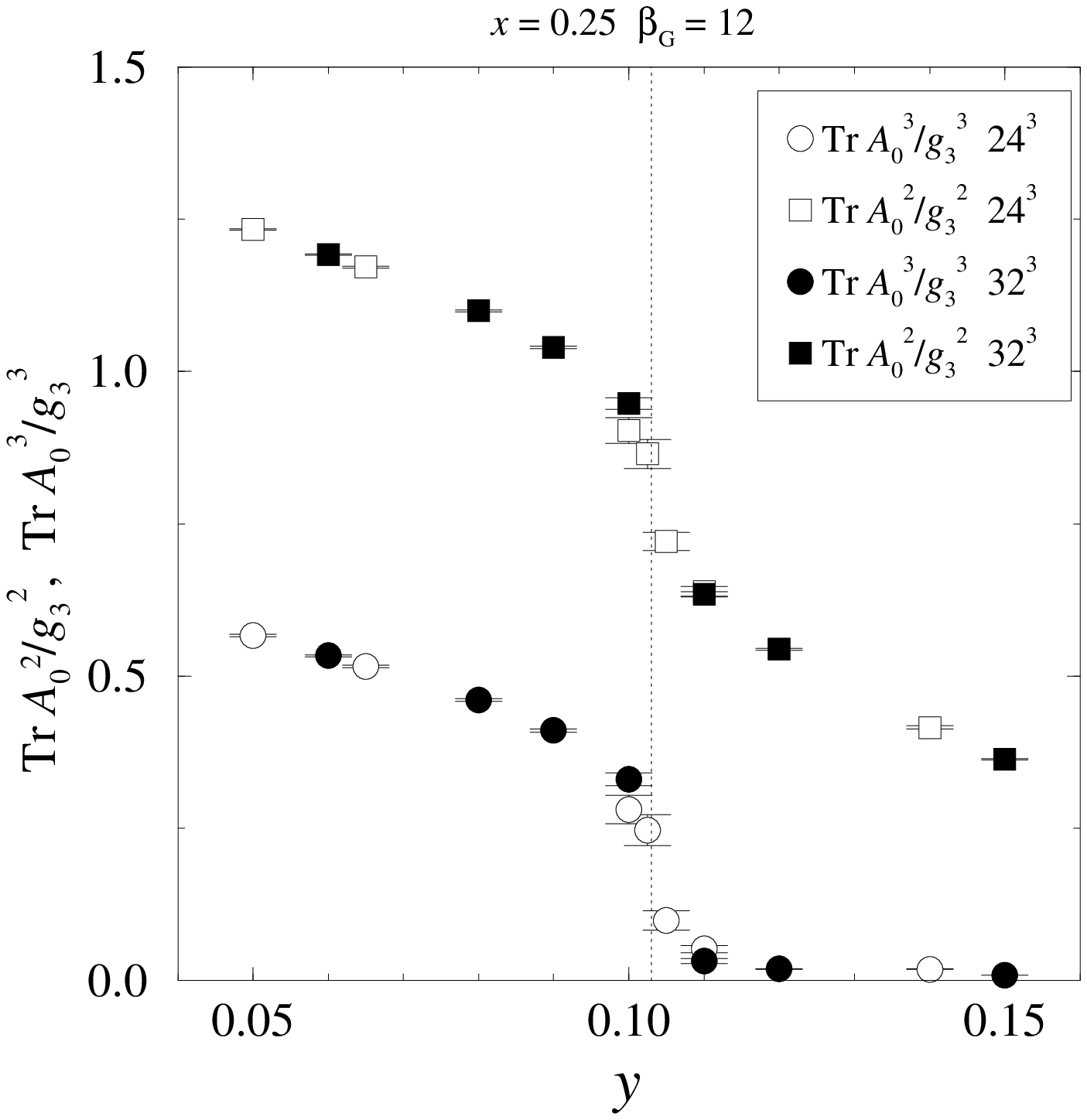} ~~~
\epsfxsize=6.8cm\epsfbox{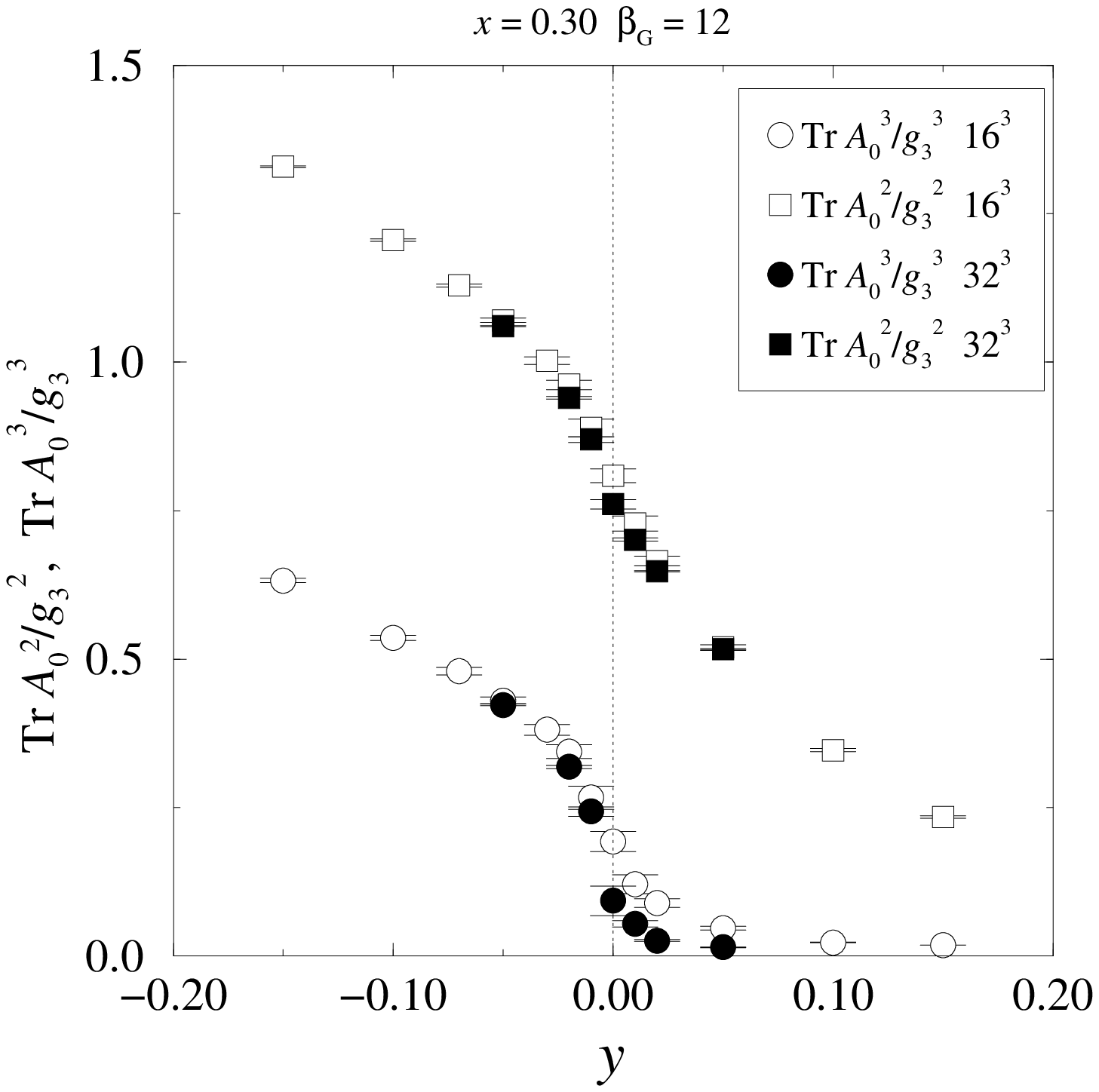}}

\caption[a]{The condensates 
$\langle \fr{1}{N^3} \sum_x \ttwo/g_3^2 \rangle$ and 
$\langle | \fr{1}{N^3} \sum_x \tthree/g_3^3 | \rangle$ measured
across the transition at $x=0.15$, 0.20, 0.25 and 0.30
(converted to continuum 
units according to \eqs\nr{tra02}, \nr{tra03}). 
The broken
phase values compare quite well with the 1-loop perturbative ones
in \eqs\nr{pert2}, \nr{pert3} for $x=0.15$. }
\la{fig:condensates}
\end{figure}

In \fig\ref{fig:condensates} we show the behaviour of the
dim-1 and dim-(3/2) condensates (scaled by a proper power of
$g_3$ to make them dimensionless),
\be
  \langle \fr{1}{N^3} \sum_x \ttwo/g_3^2 \rangle
 ~~~~~\mbox{and}~~~~~ 
  \langle | \fr{1}{N^3} \sum_x \tthree/g_3^3 | \rangle, \la{measured}
\ee
at $0.15 \le x \le 0.30$ across the transition.  Note that in the
latter case a projection to positive values is needed,
since otherwise the operator would always yield zero when measured
from a finite volume.  In the infinite volume limit, 
$\langle | \fr{1}{N^3} \sum_x \tthree/g_3^3 | \rangle$ 
is perturbatively given by \eq\nr{pert3}.
In the figures, the 
lattice operators in \eq\nr{measured} are 
converted to continuum units
according to \eqs\nr{tra02}, \nr{tra03}.

\begin{figure}[tb]
\centerline{ 
\epsfxsize=6.5cm\epsfbox{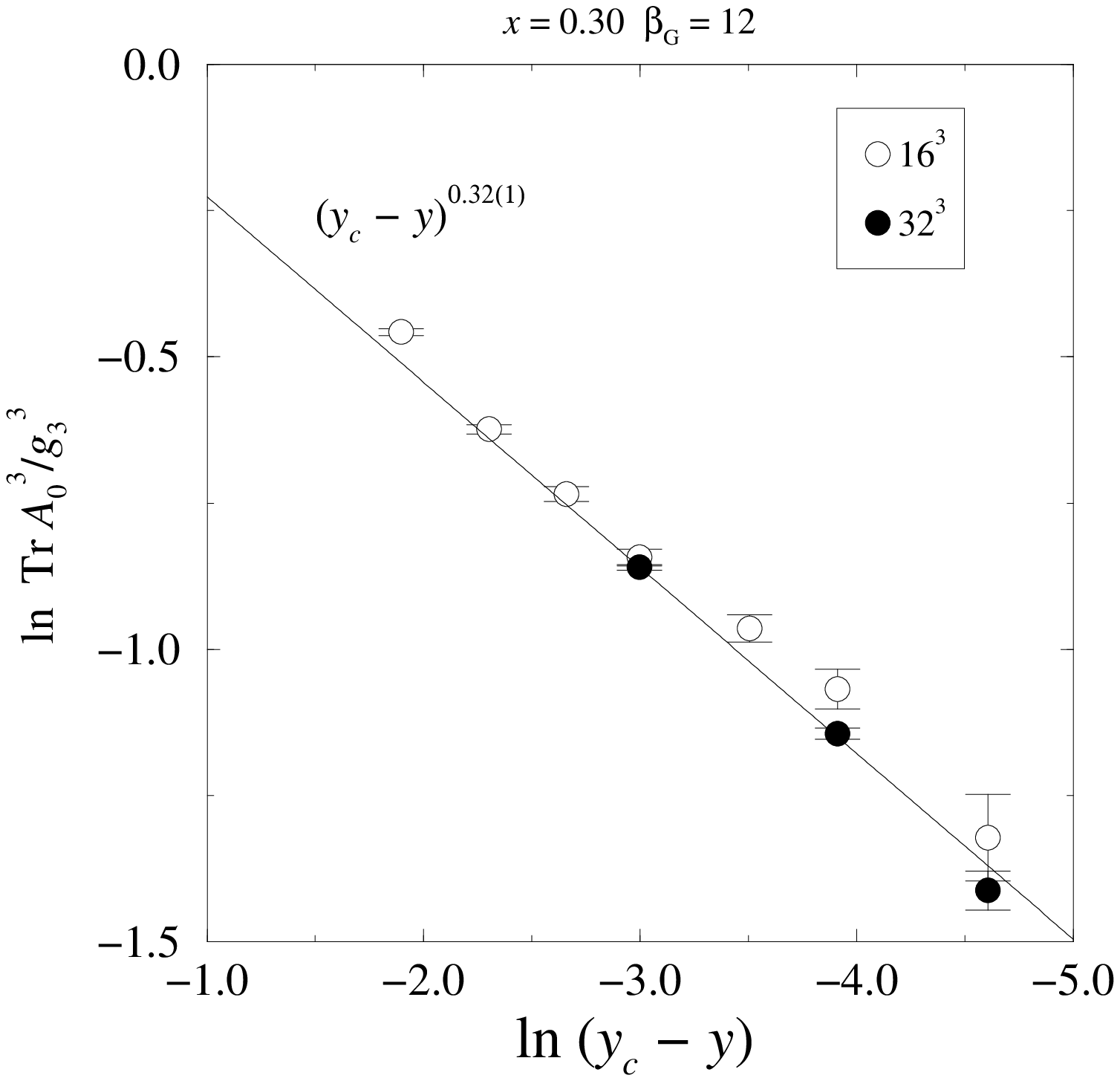} ~~~
\epsfxsize=6.5cm\epsfbox{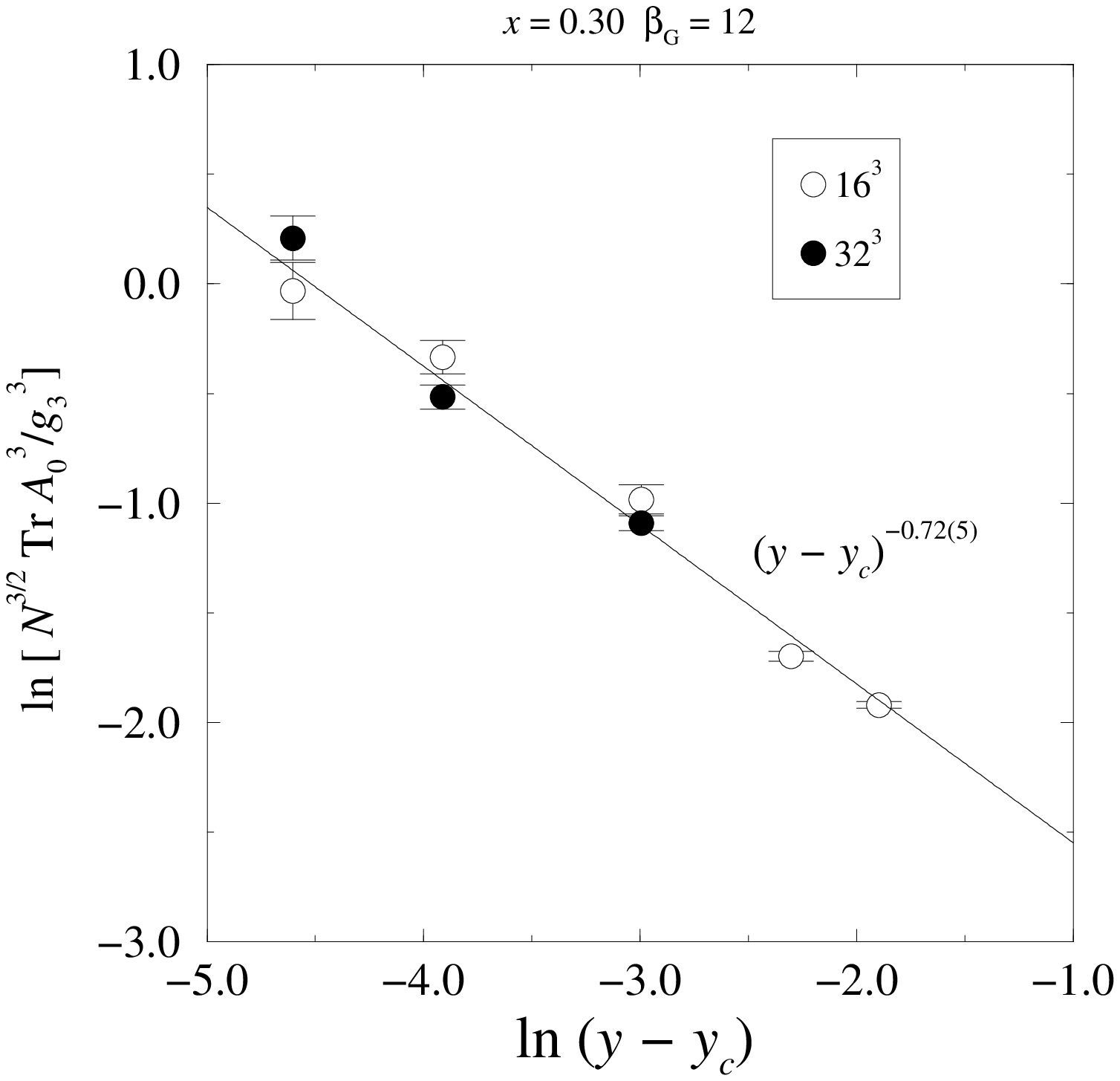}}
\caption[a]{
The scaling behaviour of $\langle | \fr{1}{N^3} \sum_x \tthree/g_3^3 |
\rangle$ for $y<y_c$ (left), $y>y_c$ (right), at finite volumes and in
a suitable range of $y$. The data is the same as in \fig\ref{fig:condensates}.
The Ising model exponents are $\approx 0.33$
for $y<y_c$ and $\approx -0.62$ for $y>y_c$ (see, e.g.,~\cite{gzj}). 
The continuous lines are fits to the $32^3$ data.\la{scaling}}
\end{figure}

At $x\le 0.2$ the transition is strongly of the 1st order, and there is a
substantial meta\-stability range across the transition point $y_c$.
However, the operator $\langle | \fr{1}{N^3} \sum_x \tthree/g_3^3 |
\rangle$, which has no additive renormalization, remains a good order
parameter at all values of~$x$, even when the metastability disappears
at $x\gsim 0.25$. Its deviation from zero at $y>y_c$ is a well
understood finite volume effect: 
for large enough volumes, it is proportional to $1/\sqrt{\mbox{volume}}$.

At $x=0.30$ we are substantially away from the first order region, and
at $y=y_c$ there is a second order transition separating the two
$\langle\tthree\rangle \ne 0$ phases from the $\langle\tthree\rangle =
0$ phase.  Indeed, $\tthree$ behaves exactly as magnetization in the
3d Ising model (this can be understood by comparing the regime
$x>x_\rmi{tricritical}$ of the phase diagram in the left panel in
\fig\ref{fig:tricritical} with that of the Ising model).  We verify
the Ising type of the transition by examining the critical behaviour of
$\langle | \fr{1}{N^3} \sum_x \tthree/g_3^3 |
\rangle$ in the neighbourhood of $y_c$.  For $y<y_c$
and in the infinite volume limit it approaches zero with the critical
exponent $\beta \approx 0.33$:
\be
  \langle | \fr{1}{N^3} \sum_x \tthree/g_3^3 | 
  \rangle  \propto (y_c - y)^\beta.
  \la{crittwo}
\ee
When $y>y_c$, $\langle | \fr{1}{N^3} \sum_x \tthree/g_3^3 |
\rangle = 0$ in the infinite volume limit.
However, at a finite volume, its behaviour
can be approximated in some range of $y$ 
(where the linear extension of the lattice is still much 
larger than the longest correlation length) as 
\be
\langle | \fr{1}{N^3} \sum_x \tthree/g_3^3 | \rangle  \propto
N^{-3/2} (y - y_c)^{-\gamma/2}\,, \la{critthree}
\ee 
where for the 3d Ising model $\gamma \approx 1.24$.
As can be seen from \fig\ref{scaling}, the data agree rather well 
with these Ising model expectations.

\begin{figure}[tb]
\centerline{ 
\epsfxsize=9cm\epsfbox{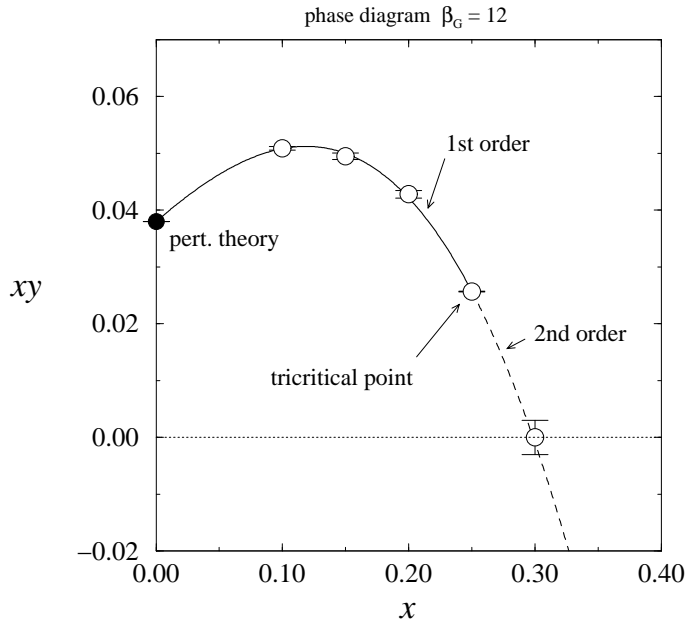}}
\caption[a]{The phase diagram of the 3d SU(3) + adjoint Higgs
theory.  The open symbols are results from the simulations, and
the filled circle is the perturbative result in \eq\nr{crline}.  The
transition line is a polynomial fit to the data.}
\la{fig:phasediag}
\end{figure}

The measured $(x,y)$-plane phase diagram is shown in
\fig\ref{fig:phasediag}.  The first order line is converted into a
second order transition at a tricritical point, which is approximately
located at $x_c \approx0.26$, $y_c \approx 0.104$
($x_\rmi{{\it c}, improved}\approx 0.24$
according to \eq\nr{ximp}). The fact that 
there is a second order transition at $x>x_c$, is based on the 
observations that (1) the transition is not of the first order
since the local order parameters behave continuously 
(\fig\ref{fig:condensates}), (2) there is a real symmetry $A_0\to -A_0$
which gets broken, and shows the scaling expected (\fig\ref{scaling}), 
(3) the correlation length related to $A_0$ seems to diverge 
at the transition point (\se\ref{sec:cor}).

The critical exponents associated with the tricritical point in 3d
assume their mean field values.  We did not attempt to
numerically analyze the critical properties at the tricritical point,
since a meaningful analysis would have required simulations with
much larger volumes and higher statistics than those used in this
work.

\subsection{Correlation lengths}
\la{sec:cor}

\begin{figure}[tb]
\centerline{ 
\epsfxsize=6.5cm\epsfbox{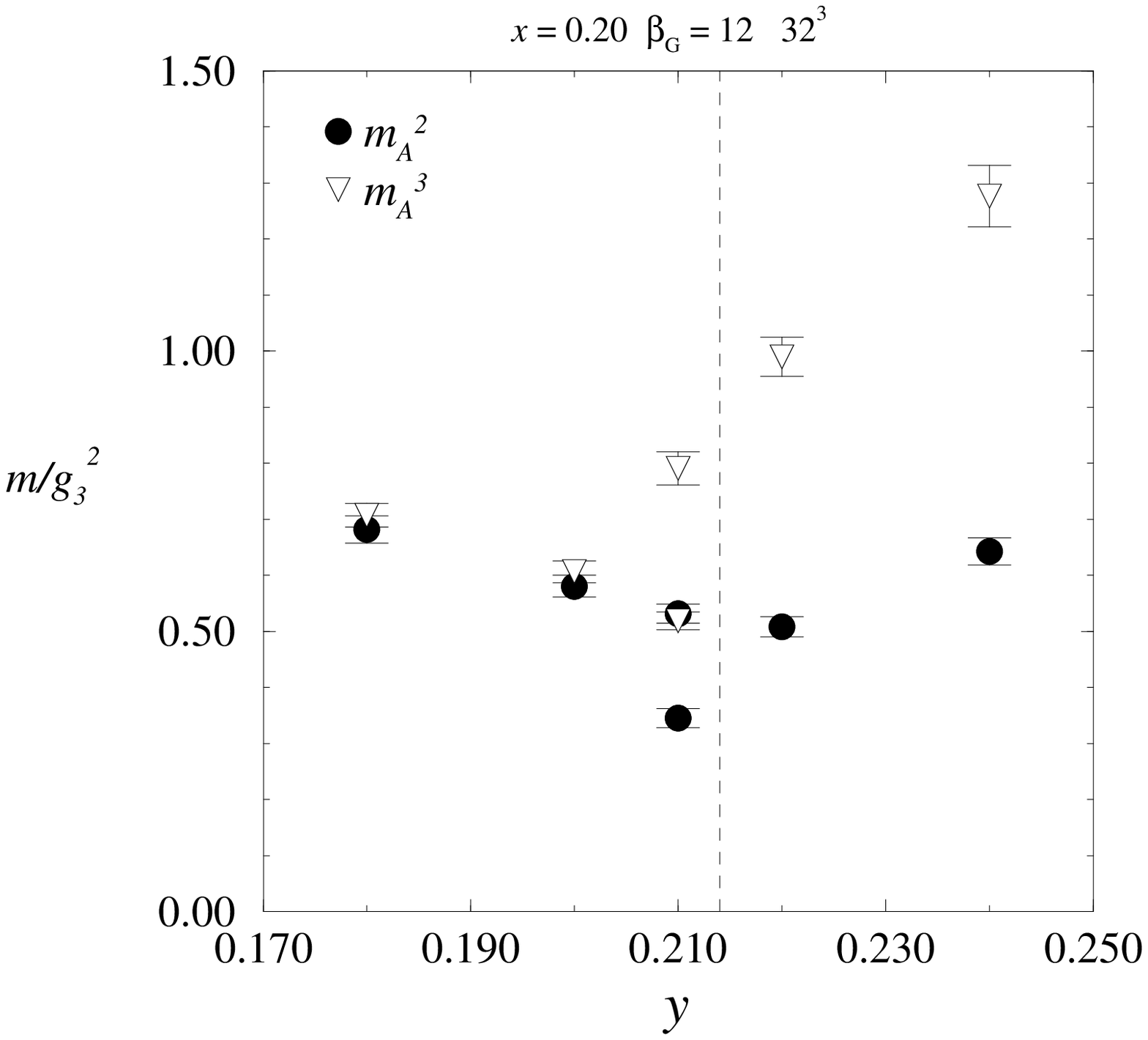} ~~~
\epsfxsize=6.5cm\epsfbox{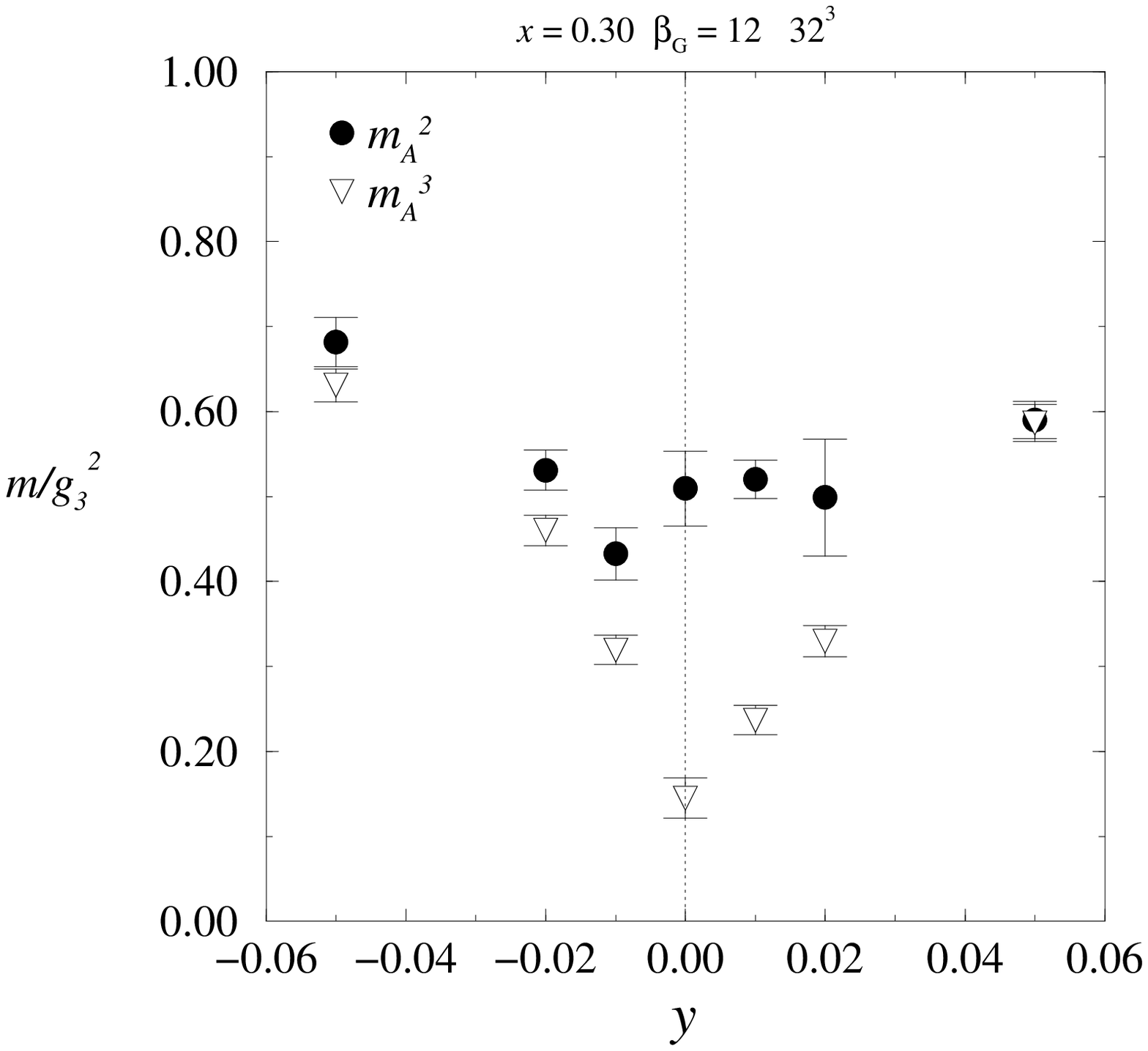}}
\caption[a]{The screening masses of the scalar operators $\ttwo$ and $\tthree$
at $x=0.2$ and 0.3. In the broken phase, both operators couple to the 
same states.\la{fig:mh}}
\end{figure}

\begin{figure}[tb]
\centerline{ 
\epsfxsize=6.8cm\epsfbox{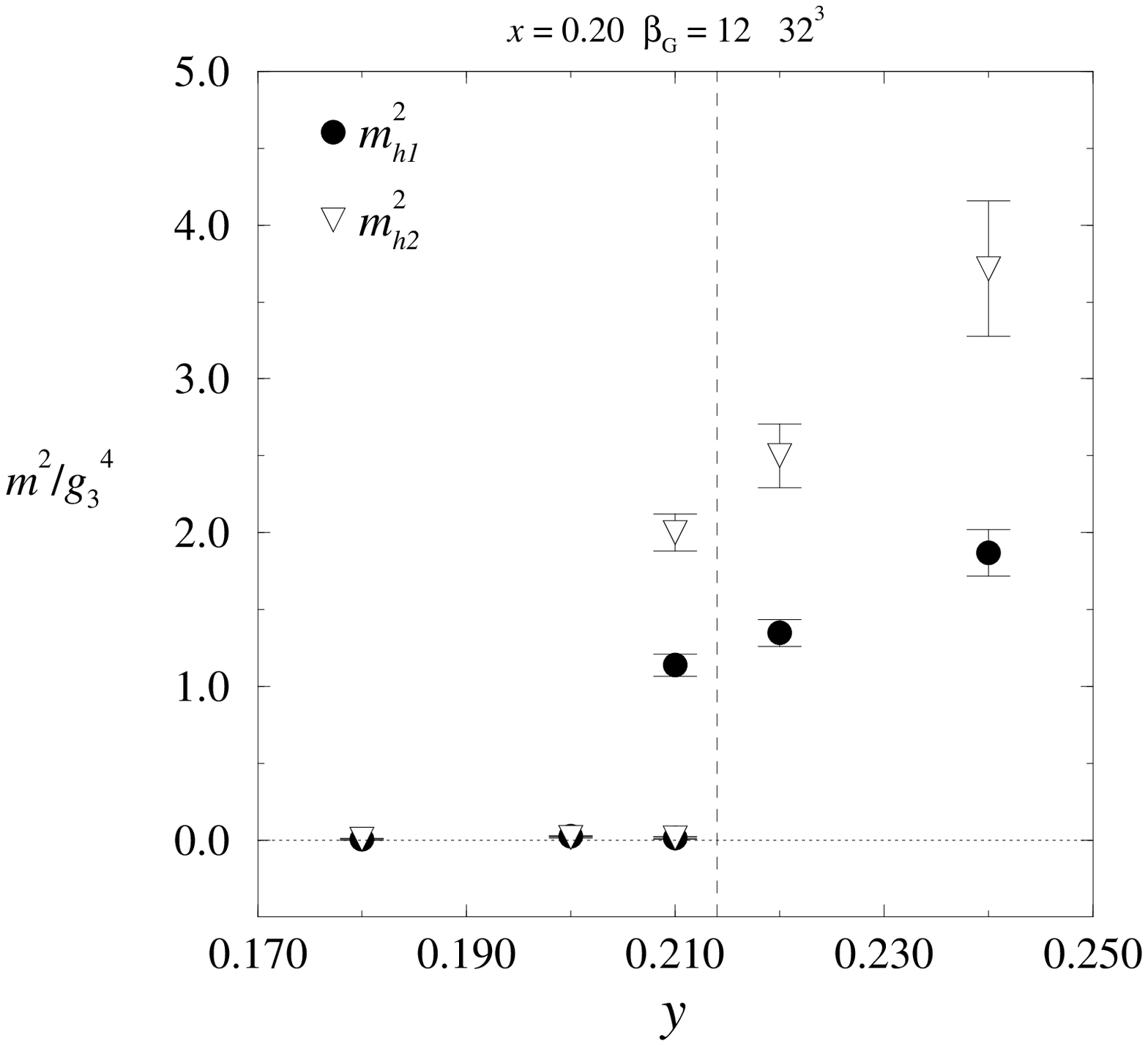} ~~~
\epsfxsize=6.8cm\epsfbox{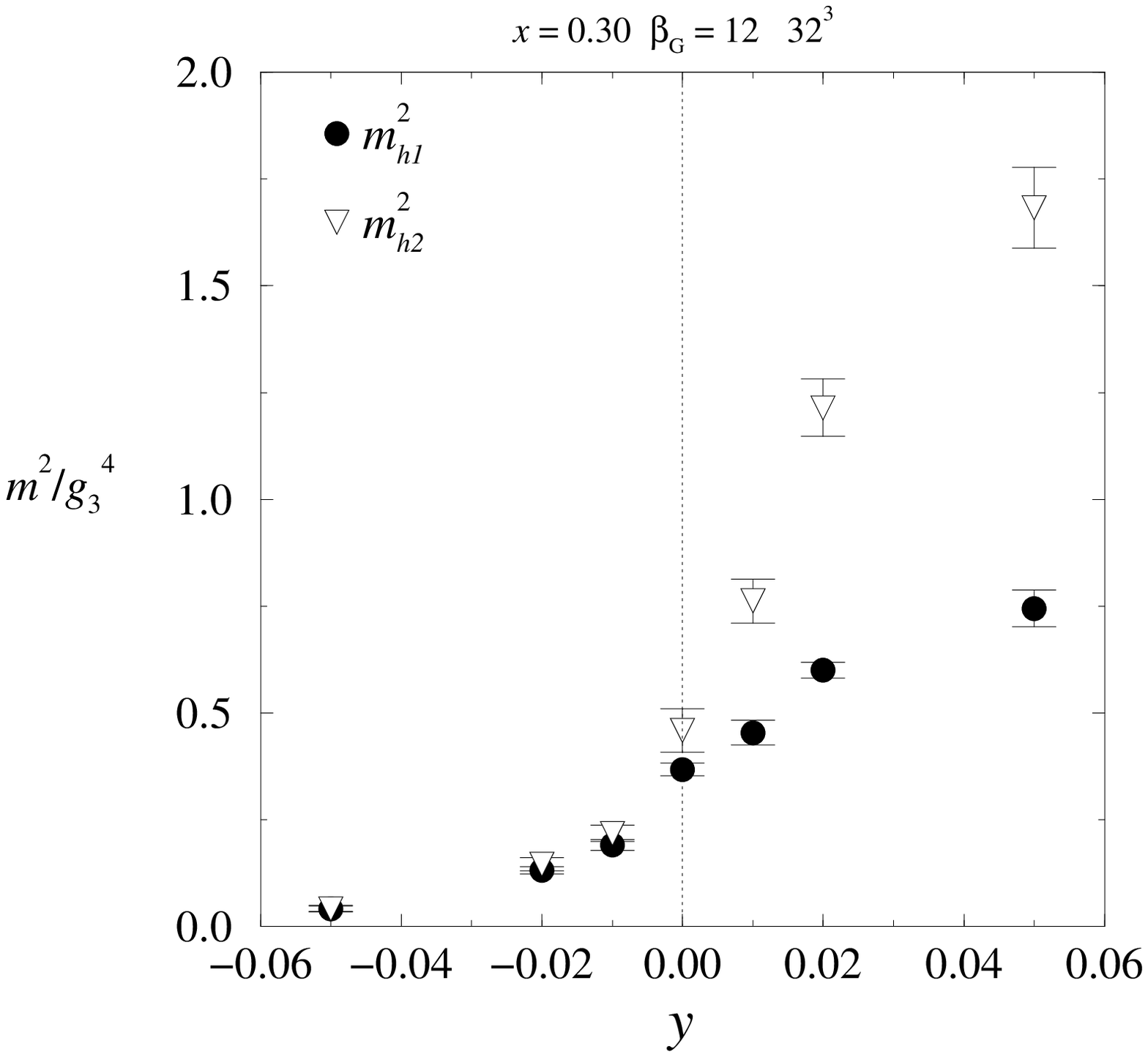}}

\caption[a]{The screening masses squared of the vector operators 
$h_{1} = \tr A_0 F_{12}$ and 
$h_{2} = \tr A^2_0 F_{12}$ at  $x=0.2$ and 0.3. 
In the broken phase, both operators couple to the 
same states.
\la{fig:m2h}}
\end{figure}

The spatial correlation lengths, or inverse
screening masses, were measured with a method using recursive
levels of smearing and blocking of the gauge and $A_0$ variables.  We
used up to 4 blocking levels, which gives $2^4$ as the largest spatial
extent of the operators.  For technical details, we refer 
to~\cite{su2adjoint}.  

In Figs.~\ref{fig:mh} and \ref{fig:m2h} we show the screening masses
of the scalar operators $\ttwo$ and $\tthree$ and the vector operators
\be 
 h_{1,i} = \epsilon_{ijk}\tr A_0 F_{jk} 
\mbox{~~~~and~~~~}
 h_{2,i} = \epsilon_{ijk}\tr A^2_0 F_{jk}
\ee
at $x=0.20$ and 0.30.  In the symmetric phase, 
$\ttwo$, $\tthree$ have different quantum 
numbers and thus couple to different states; 
the same is true for $h_1,h_2$. However, in the 
broken $A_0$ phase the operators can couple to
each other, and should thus project to the same 
states. This is indeed observed in \fig\ref{fig:mh} 
within the statistical errors for $x=0.20$. For $x=0.30$
the signal of the $\ttwo$ correlator is quite noisy
(and the errorbars do not contain any estimates of the
systematic effects), but the pattern should be the same. 
Note that this does not as such imply that there would be 
a discontinuity in the mass spectrum at $y=y_c,x=0.30$:
it is just that in the broken phase $\ttwo,\tthree$
couple to the same excitations, and to determine the mass
of the first exited state would require 
a mixing analysis, such as in~\cite{ptw}. 
In the symmetric phase, in contrast, $\ttwo,\tthree$
automatically couple to different states. 

The scalar operator $\tthree$ becomes
``critical'' at $x=0.30$: it approaches zero at $y=y_c$, as much as
allowed by the finite volume.  This is in accordance with the
interpretation that $\tthree$ acts like the magnetization of 
the 3d Ising model.  A precise finite size scaling analysis of 
the critical correlation length is beyond the scope of this paper,
and we rather interpret the right panel of \fig\ref{fig:mh}
as a consistency check. 

In perturbation theory, there always remains an
unbroken U(1) gauge symmetry in the broken phase.  
The vector operators $h_1$ and $h_2$
couple to the massless U(1) ``photons''~$\gamma$.  However, these
photons become massive due to the interactions with Polyakov monopoles.
Since we expect the screening mass to be very light, we measure it
at non-zero transverse momentum: we use the operator
\be
   O_1(x_3) = \fr{1}{N^2}\sum_{x_1,x_2} \tr A_0 F_{12} e^{i2\pi x_1/N}
\ee
and correspondingly for $h_2$
(on the lattice, $\tr A_0 F_{12}\to\tr A_0 U_{12}$).  
The screening mass is then measured
from the asymptotic behaviour of the correlation function,
\be
  \langle O_1(x_3)O_1(0)\rangle \propto
  \exp\Bigl(-x_3 \sqrt{(2\pi/N)^2 + m^2_{h_1} } \Bigr)\,.
\ee
The results are shown in \fig\ref{fig:m2h}.  When $x=0.2$, the
transition is strong and, within the statistical accuracy, the masses
fall to zero in the broken phase.  However, at $x=0.3$ the masses
remain finite even in the broken phase.  Due to the analytic
continuation this implies that also at $x=0.2$ the masses are still
finite; they are just too small to be observed.

It is interesting to note that the vector masses seem to vary rather
smoothly when going from the symmetric to the broken phase, despite 
the fact that there is a true symmetry breaking phase transition.
In fact, their behaviour is quite similar to that in the 
SU(2) + adjoint Higgs theory~\cite{su2adjoint}, 
where no transition at all is 
observed at large~$x$~\cite{hpst}. 

\section{Conclusions}
\la{sec:con}

We have numerically determined the phase diagram of the 
3d SU(3) + adjoint Higgs theory. We find a symmetric phase
with $\langle\ttwo\rangle\approx 0$, 
$\langle\tthree\rangle =0$, and a broken phase
with $\langle\ttwo\rangle, \langle\tthree\rangle$ 
non-vanishing and large. These are separated by a transition 
line which contains a first order regime, a second order
regime, and a tricritical point in between. We did not
observe any other types of phases. 

{}From the statistical physics point of view, these results
are of interest as a new qualitative class in the types of
critical behaviours that have been found in 3d SU($N$)+Higgs 
theories. (It should be noted that even more classes could
appear when external constraints such as a magnetic field 
are added, or when the Higgses are replaced by fermions~\cite{fmm} --
the latter case is not relevant for finite temperature 4d theories, 
though. It can also be noted that a similar tricritical structure
as found here can arise when there are several (fundamental) Higgses
in the theory, related by a global symmetry~\cite{aw}.) 
It would be interesting to apply methods similar
to those in~\cite{endpoint} to a more precise study of
the properties of the tricritical point. To our knowledge,
the universal forms of the different  
probability distributions at the tricritical point 
have not been previously determined numerically in 
three-dimensional theories. 

{}From the QCD point of view, we re-emphasize the fact that 
even though the SU(3)~+ adjoint Higgs theory has the phase structure
given in this paper, only the symmetric phase
is an accurate effective theory for the 4d finite temperature theory,
permitting, say, the determination of static correlation functions. 
In particular, the phase transition of the
effective theory is not that of the 4d theory. 
(Nevertheless,
it is amusing to note that the location of the tricritical point
$x_c$ corresponds to $T\approx0.57\lambdamsbar$ according
to 2-loop dimensional reduction, not far from the 4d transition
temperature $T_c\approx1.03\lambdamsbar$ \cite{fingberg}.)
One may also note that in the context of finite $T$ QCD
it has been suggested~\cite{qcd}
that the critical line in the $(T,\mu)$-plane for two massless 
flavours would have a similar tricritical
structure. This clearly is not related to the tricritical
structure studied here.

Finally, from the point of view of the finite temperature SU(5) 
theory, the present SU(3) case shows that only some of the 
possible broken phases are dynamically realized. It should 
be interesting to see how this statement is modified in the
SU(5) case, where there is a wider spectrum of possibilities.

\section*{Acknowledgements}

Most of the simulations were carried out with a Cray T3E 
at the Center for Scientific Computing, Finland.  We thank 
G.D. Moore and M. Shaposhnikov for useful discussions. This work was
partly supported by the TMR network {\em Finite Temperature Phase
Transitions in Particle Physics}, EU contract no.\ FMRX-CT97-0122.
The work of A.R. was partly supported by the University of Helsinki, 
and the work of M.T. by the Russian Foundation
for Basic Research, Grants 96-02-17230, 16670, 16347.

\end{document}